\documentclass{achemso}

\usepackage{amssymb}
\usepackage{amsmath} 
\usepackage{threeparttable}

\newcommand{\parallelsum}{\mathbin{\!/\mkern-5mu/\!}}
\SectionNumbersOff

\author{F. Le Roux}
\email{florian.leroux@physics.ox.ac.uk}
\affiliation[University of Oxford]
{Department of Physics, University of Oxford, Parks Road, Oxford, OX1 3PU, U.K.}

\author{R. A. Taylor}
\affiliation[University of Oxford]
{Department of Physics, University of Oxford, Parks Road, Oxford, OX1 3PU, U.K.}

\author{D. D. C. Bradley}
\alsoaffiliation[KAUST]
{Physical Science and Engineering Division, King Abdullah University of Science and Technology, Thuwal, 23955-6900, Saudi Arabia}

\title{Exciton-Polaritons in Uniaxially Aligned Organic Microcavities}


\begin{document}

\begin{abstract}

Here we report the fabrication and optical characterization of organic microcavities containing liquid-crystalline conjugated polymers (LCCPs): poly(9,9-dioctylfluorene-co-benzothiadiazole) (F8BT), poly(9,9-dioctylfluorene) (PFO) and poly(2,7-(9,9-dihexyl \newline fluorene)-co-bithiophene) (F6T2) aligned on top of a thin transparent Sulfuric Dye 1 (SD1) photoalignment layer. We extract the optical constants of the aligned films using variable angle spectroscopic ellipsometry and fabricate metallic microcavities in which the ultrastrong coupling regime is manifest both for the aligned and non-aligned LCCPs. Transition dipole moment alignment enables a systematic increase in the interaction strength, with unprecedented solid-state Rabi splitting energies up to 1.80 eV for F6T2, the first to reach energies comparable to those in the visible spectrum; with an optical gap of 2.79 eV this also gives the highest-to-date organic microcavity coupling ratio, 65$\%$. We also demonstrate that the coupling strength is polarization-dependent with bright polaritons photoluminescence for TE polarization parallel to the transition dipoles and either no emission or weakly coupled emission from the corresponding TM polarization. The use of uniaxally aligned organic microcavities with switchable coupling strength offers exciting prospects for direct observations of ultrastrong coupling signatures, quantum simulation, polaritonics and condensation related phenomena.

\end{abstract}

\newpage

The strong coupling (SC) regime in the solid-state is entered\cite{Weisbuch1992} when the interaction between the electric component of a confined electromagnetic field and the excitations present within a semiconductor becomes sufficiently intense that their original energy levels are replaced by so-called polariton hybrid states of light and matter, separated by a Rabi splitting energy $\hbar\Omega_{\rm R}$. Organic semiconductor Frenkel excitons are an interesting alternative to the more traditional Wannier excitons seen in III-V inorganic semiconductors for the study of exciton-polaritons thanks to their large binding energies ($E_{B}\sim0.5\pm0.25\;$eV\cite{Marks1994,Alvarado1998}) which allow room-temperature observation of varied phenomena including Bose-Einstein condensation\cite{Plumhof2014,Daskalakis2014}, superfluidity of light\cite{Lerario2017} and optical logic\cite{Zasedatelev2019}. Their large intrinsic oscillator strengths\cite{Lidzey1998} combined with the small mode volumes $V_{\rm m}$ of metallic microcavities\cite{Schwartz2011} have enabled $\hbar\Omega_{\rm R} \geq 1\;$ eV\cite{Kena-Cohen2013,Gambino2014,Mazzeo2014,Suzuki2019}, with values up to $\hbar\Omega_{\rm R} = 1.12\;$eV \cite{Liu2015,Liu2019}. This splitting is directly comparable to the exciton transition energy $\hbar\omega_{\rm ex}$ an yields normalized coupling ratios $g = \frac{\Omega_{\rm R}}{\omega_{\rm ex}}\geq20\%$, thereby crossing into ultrastrong coupling (USC), an interaction space that has received great recent attention, with attractive research perspectives and multiple emerging applications\cite{FriskKockum2019,Forn-Diaz2019}. Experimental realizations of increasingly higher coupling ratios have also been reported for inorganic semiconductor based intersubband polaritons\cite{Askenazi2017}, and other physical systems, including superconducting circuits\cite{Yoshihara2018}, Landau polaritons\cite{Bayer2017} and plasmonic picocavities interacting with vibrational degrees of freedom of individual molecules\cite{Benz2016}. 

For an ensemble of organic semiconductor excitons within a cavity, $\hbar\Omega_{\rm R}$ scales with the square root of $\omega_{\rm ex}$ according to \cite{Ciuti2005,George2015,Tropf2018}: 

\begin{equation}
\label{eq:Scaling}
\hbar\Omega_{\rm R} = 2\boldsymbol{{\mu}.E}\sqrt{\frac{N\hbar\omega_{\rm ex}}{2\epsilon_{\rm eff}V_{\rm m}}},
\end{equation}

where $\boldsymbol{{\mu}}$ is the transition dipole moment, $\boldsymbol{E}$ the electric field, $N$ the number of molecules, $\hbar\omega_{\rm ex}$ the exciton transition energy, $\epsilon_{\rm eff}$ the cavity effective permittivity and $V_{\rm m}$ the cavity mode volume.

One way to increase the value of the coupling ratio has then been to work with lower energy excitons, as done by Barachati et al.\cite{Barachati2018}, resulting in a then record $g = 62\%$. This approach is inherently accompanied by a reduction of $\hbar\Omega_{\rm R}$ compared with the use of excitons lying at higher energies with equivalent oscillator strengths. The alternative is to look to increase $\hbar\Omega_{\rm R}$; The most direct routes to achieve this include (i) increasing $N$, which although generally not straightforward can be done, for example, by reducing the bulkiness of conjugated polymer solubilizing groups\cite{Campoy-Quiles2008} (ii) increasing $\boldsymbol{{\mu}}$ through conformational control\cite{LeRoux2018,Perevedentsev2016} or a photo-switchable configuration change\cite{Schwartz2011}, and/or (iii) increasing $\boldsymbol{{\mu}.E}$. In the latter case, uniaxial orientation has been shown to enhance conjugated polymer thin film refractive index (and correspondingly transition dipole moment) in the direction parallel to the chain orientation axis\cite{Virgili2001,Campoy-Quiles2005b} yielding an enhanced dot product for a suitable polarization of $\boldsymbol{E}$. This offers a clear route to enhancing $\hbar\Omega_{\rm R}$ that is demonstrated below, using a photoalignment process to achieve thermotropic liquid crystalline conjugated polymer (LCCP) chain orientation. Two recent reports on the coupling of liquid crystal (LC) vibrational modes\cite{Hertzog2017} and carbon nanotubes Wannier excitons\cite{Gao2018} have also shown that $\boldsymbol{{\mu}.E}$ can be maximized in this way and that polarization-dependence allows for applications discussed further in the text.

Previous approaches to LCCP orientation typically used a traditional rubbed polyimide (PI) alignment layer onto which the polymer was spin coated prior to thermal treatment\cite{Grell1997}. The clearing temperatures of LCCPs are relatively high ($\sim$200-300$^{\circ}$C)\cite{Grell1997,GRELL1999}, leading to the requirement for a high temperature stable PI, for which there are limited commercial options. Precursor route poly(p-phenylenevinylene) has also been used as an alternative rubbed alignment layer, having the advantage of temperature stability and an electronic structure that more readily permits charge injection from the underlying electrode to the LCCP\cite{Whitehead2000a}. Other approaches to orientation include stretching and rubbing the conjugated polymer\cite{Dyreklev1995}, Langmuir-Blodgett deposition\cite{Cimrova1996}, use of an aligned host matrix\cite{Hagler1991} or nanoimpriting\cite{ZijianZheng2007}. Successful fabrication of polarized light emitting diodes\cite{Whitehead2000,Schmid2008,Lee2017,Whitehead2000,Whitehead2000a,Misaki2008} and polarized photoluminescence structures\cite{Virgili2001} has resulted, together with intrachain mobility enhanced transistors\cite{Sirringhaus2000}. For a variety of practical reasons, oriented LCCPs have, however, not been used before in strongly or ultrastrongly-coupled microcavities. 

Non-contact photoalignment of LC mesophases has emerged\cite{Seki2014} as a promising alternative to rubbing-induced alignment. Among photoalignment layer materials, the azobenzene-containing Sulfuric Dye 1 (SD1) has shown high temperature stability and remarkable quality for the alignment of low molecular weight LCs\cite{Li2006,Ma2015}. Recently, the orientation of poly(9,9-dioctylfluorene-co-benzothiadiazole) (F8BT) using SD1 has also been observed\cite{Zhang2019} in which a thermal treatment enables the orientation of the LCCP. Potential advantages of using SD1 photoalignment for photonics and polaritonics are threefold, namely that the SD1 layer can be very thin ($\leq 5\;$ nm), that it is almost transparent in the visible (peak absorption at $\sim$ 3.25 eV) and that it is a patternable process which allows a straightforward way to fabricate novel photonic structures\cite{Ma2015,Ma2017}.

We report here a detailed study of the use of oriented LCCP films within metallic microcavities. F8BT, poly(9,9-dioctylfluorene) (PFO) and poly(2,7-(9,9-dihexylfluorene)-co-bithiophene) (F6T2) films are oriented with SD1 photoalignment layers and in all three cases we demonstrate a systematic enhancement of $\hbar\Omega_{\rm R_{\rm TE}}$, for TE-polarized light parallel to the chain orientation direction, compared to non-aligned reference samples. The maximum $\hbar\Omega_{\rm R_{\rm TE}}= 1.80\pm 0.01\;$eV (689 nm) is for F6T2, a value that would sit within the visible spectrum. This structure also gives the largest normalized coupling ratio, $g = 65 \%$, reported to date for an organic semiconductor microcavity. Photoluminescence for TE-(parallel to the alignment direction) and corresponding TM-polarizations makes the changes in coupling strength between polarizations evident for all three polymers. Going beyond enhancement of the Rabi-splitting energy, we  discuss the potential use of uniaxially aligned organic microcavities for demonstration of the elusive polaritonic NOT gate\cite{Sanvitto2016,Solnyshkov2015,Espinosa-Ortega2013}, for quantum simulation through complex energy landscapes and more generally its advantages for the realization of polarization sensitive devices, lasing and condensation related phenomena.

\section{Results And Discussion}

\begin{figure}[H]
\includegraphics[scale=0.24]{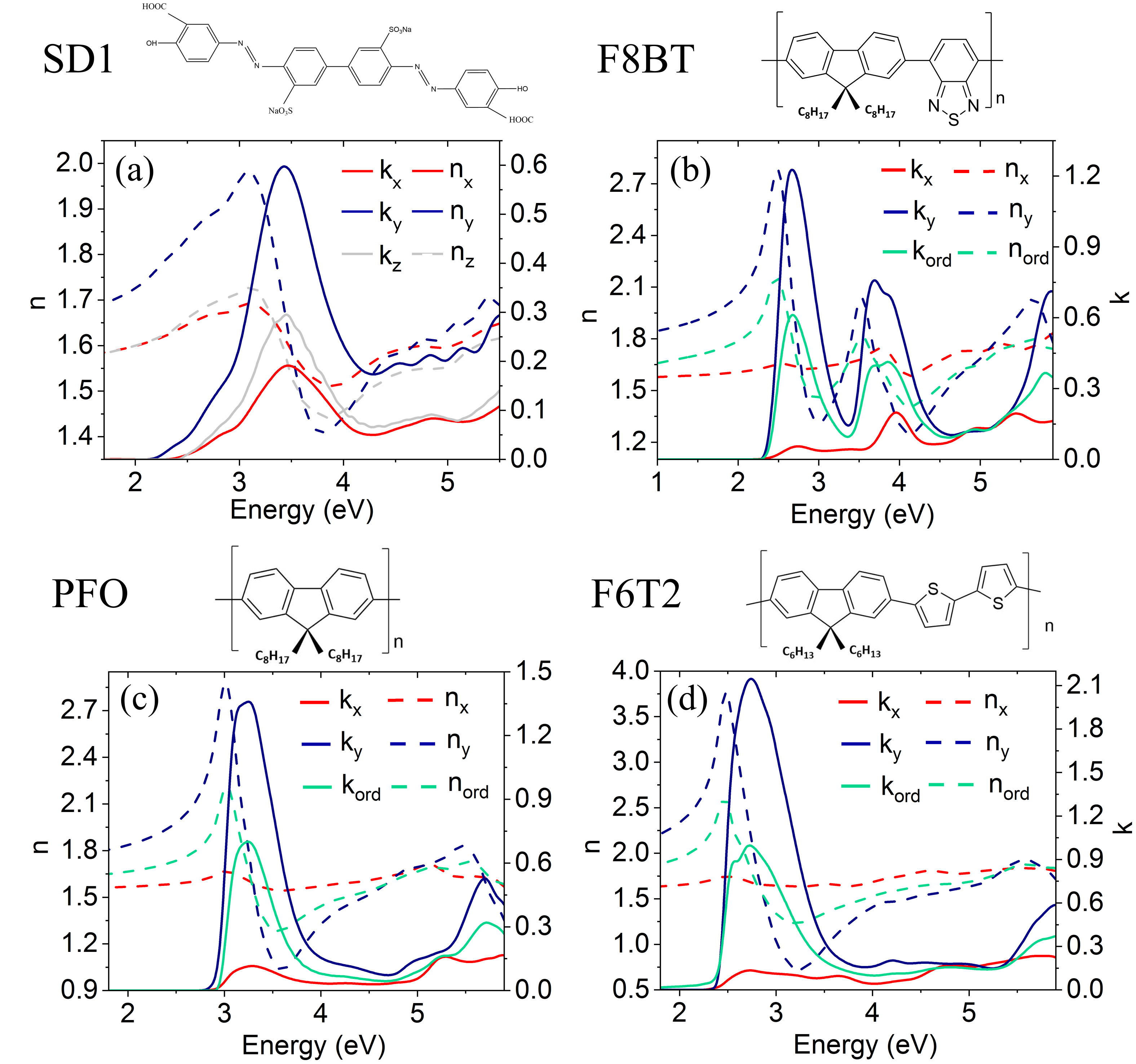}
\caption{\label{fig:FigureIndices} (a) Optical components for a thin film of SD1 aligned along the y direction. (b),(c),(d) In green: In-plane $n_{\rm ord}$, $k_{\rm ord}$ optical components for spincoated F8BT (b), PFO (c) and F6T2 (d). The in-plane optical components of the films following alignment are shown in blue for $n_{y}$, $k_{y}$ (parallel to the alignment direction) and in red for $n_{x}$, $k_{x}$ (perpendicular to the alignment direction) for SD1 (a), F8BT (b), PFO (c) and F6T2 (d). Solid lines give the real component of the complex refractive index $\tilde{n} = n + ik$, dashed lines the imaginary component.}
\end{figure}

The optical constants for thin films of PFO, F8BT, F6T2 (see Methods for all fabrication protocols) were extracted using Variable Angle Spectroscopic Ellipsometry (VASE). As the polymer chains tend to lie in the plane of the film\cite{Campoy-Quiles2005a,Campoy-Quiles2014,Campoy-Quiles2005}, the resulting optical constants are well fitted\cite{LeRoux2018,Tropf2017} using an in-plane/out-of plane anisotropic model yielding the components $n_{\rm ord}, k_{\rm ord}, n_{\rm ex}, k_{\rm ex}$ of the complex refractive index $\tilde{n} = n + ik$. Figure~\ref{fig:FigureIndices} shows (in green) the in-plane optical constants ($n_{\rm ord}, k_{\rm ord}$) for F8BT (b), PFO (c) and F6T2 (d) (The complete sets of optical components extracted are available in Supplementary Information). All spectra are comprised of either one or several inhomogeneously broadened distributions ($E_{\rm X_{\rm PFO}}$ at around $3.23\;$eV, $E_{\rm X_{\rm F8BT_{1}}}$ and $E_{\rm X_{\rm F8BT_{2}}}$ at respectively $3.82\;$eV and $2.70\;$eV and $E_{X_{F6T2}}$ at $2.79\;$eV) with excitation states lying above $5\;$eV associated to ring-localized fluorene states\cite{Rothe2006}.

Following photoalignment of an SD1 spincoated film in the in-plane y direction, the optical components $n_{x}, n_{y}, n_{z}, k_{x}, k_{y}, k_{z}$ were extracted using a biaxial anisotropic model\cite{Valyukh2008} and are shown in Figure~\ref{fig:FigureIndices}(a). As expected, the alignment brings about intense $n_{y}$, $k_{y}$ optical components compared to their perpendicular counterparts $n_{x}$, $k_{x}$. Polymer layers of F8BT, PFO and F6T2 were then spincoated on top of the photoaligned SD1, thermally annealed into their respective nematic phases ($250^{\circ}$C for F8BT, $160^{\circ}$C for PFO and $220^{\circ}$C for F6T2) and subsequently quenched to room-temperature causing and effectively freezing the in-plane uniaxial alignment in the $y$ direction. The corresponding optical components of the obtained films were fitted and are shown in Figure~\ref{fig:FigureIndices}(b),(c),(d) with $n_{y}$, $k_{y}$ substantially larger than $n_{x}$, $k_{x}$. As the oscillator strengths in the $x$ and $y$ direction $f_{x,y} \propto \int k_{x,y} (E)dE$ are proportional to the number of underlying contributing dipoles, we calculate the ratio $R= 1 - \frac{f_{x}}{f_{x} + f_{y}}$ where $f_{x,y}$ is the oscillator strength in the corresponding direction and estimate the total percentage of dipoles aligned in the $y$ direction. For all excitons, $R$ exceeds 83$\%$ (calculated for $E_{\rm X_{\rm F8BT_{1}}}$) with a maximum value of 94$\%$ (for $E_{\rm X_{\rm F6T2}}$) underlining the remarkable alignment quality.

\newpage

\begin{figure}[H]
\includegraphics[scale=0.23]{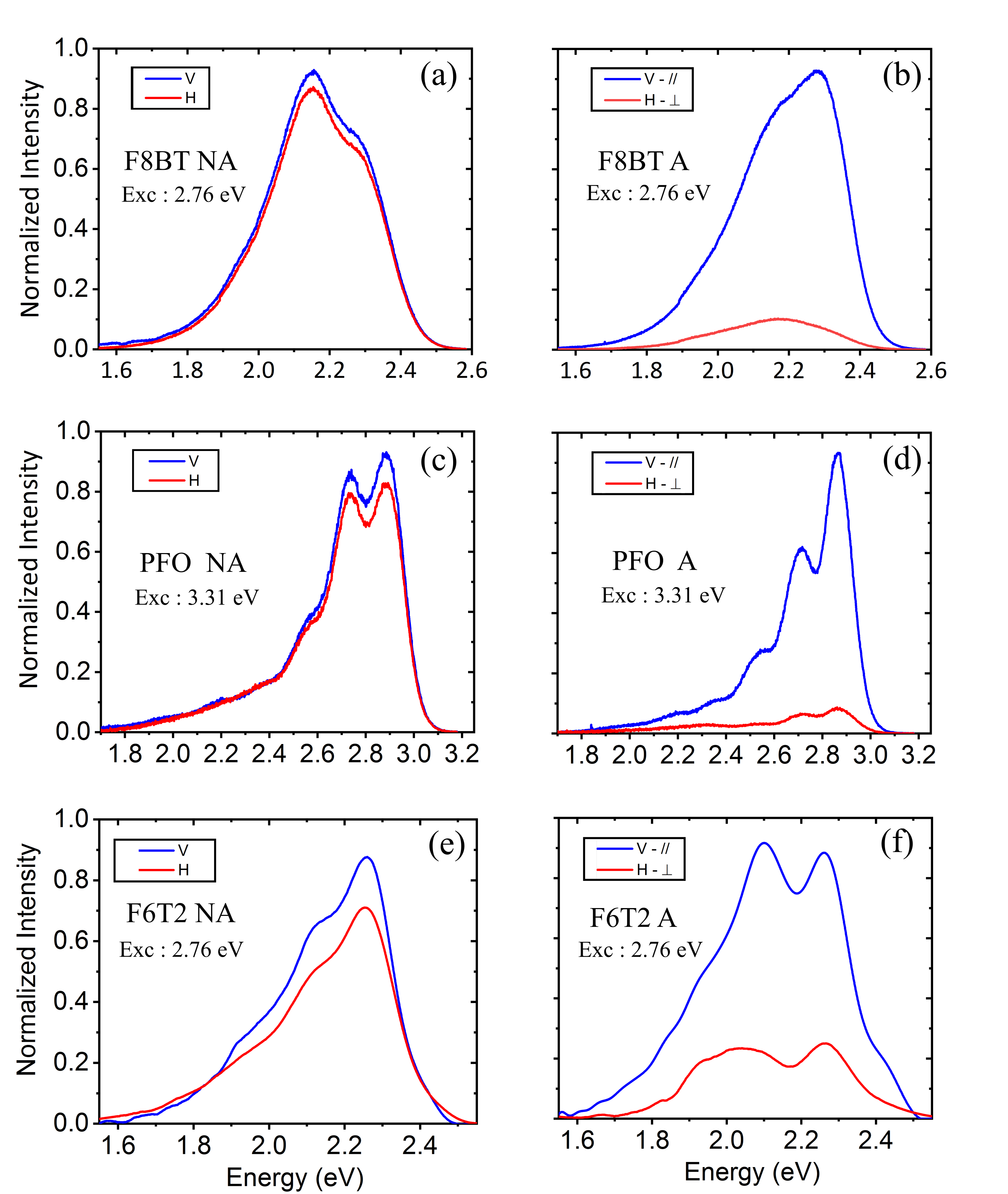}
\caption{\label{fig:FigurePLFilms} PL spectra for non-aligned ((a),(c) $\&$ (e)) and aligned ((b),(d) $\&$ (f)) F8BT ((a) $\&$ (b)), PFO ((c) $\&$ (d)) and F6T2 ((e) $\&$ (f)) thin films. For a given polymer in a given alignment state, both polarizations spectra were normalized by the same value to allow for their comparisons. Solid blue lines show the PL collected using a polarizer in the vertical direction and red lines in the horizontal direction. In panels (b),(d) and (f): $\parallelsum$ and $\perp$ indicate that the collection polarizer's direction respectively matches or is perpendicular to the direction of the polymer chains in the film. A fast Fourier transform algorithm was used for F6T2 ((e) $\&$ (f)) to reduce noise in the recorded signal.}
\end{figure}

Time-integrated photoluminescence (PL) spectra were recorded at normal incidence for the three polymers and are shown in Figure~\ref{fig:FigurePLFilms} (see Methods for the experiment geometry). The spectra were measured at normal incidence with the collection polarizer both in vertical (blue) and horizontal (red) positions; for the aligned films, the vertical direction matched the direction of the alignment. In each case, we calculate the integrated ratio $R_{\rm VH} = \int \frac{I_{\rm V}(E)}{I_{\rm H}(E)}dE$ which reveals the presence of in-plane uniaxial alignment in the film. For F8BT, the spectrum of the non-aligned film (Figure~\ref{fig:FigurePLFilms} (a)) reveals an inhomogeneously broadened distribution with $S_{1}-S_{0}$ (0-0), (0-1) vibronic peaks located at $2.29\;$eV ($541\;$nm) and $2.15\;$eV ($577\;$nm). We calculate $R_{\rm VH} = 1.08$, with a deviation from unity being fully accounted for by the degree of polarization of the excitation laser beam; the polymer chains have as expected no preferential in-plane orientation. For the aligned F8BT (\ref{fig:FigurePLFilms} (b)), the same spectral positions for the vibronic peaks are recorded with a difference in relative heights due to thickness variations between the aligned and non-aligned films. The integrated ratio $R_{\rm VH} = 8.3$ makes the in-plane preferential alignment of the emitting layer evident (in-depth studies relating the polarized emission to the microscopic parameters of aligned F8BT can be found in Ref 41). For non-aligned PFO (Figure~\ref{fig:FigurePLFilms}(c)), the $S_{1}-S_{0}$ well resolved (0-0), (0-1) and (0-2) PL vibronic peaks appear at $2.88\;$eV ($430\;$nm), $2.71\;$eV ($457\;$nm) and $2.58\;$eV ($481\;$nm) with an integrated ratio $R_{\rm VH}$ that increases from 1.09 to 6.9 from non-aligned to aligned film (Figure~\ref{fig:FigurePLFilms}(d)). As the vibronic structure is this time well resolved, we calculate the peak ratios $R_{\rm pVH}(E) = \frac{I_{\rm V}(E)}{I_{\rm H}(E)}$ for $R_{\rm pVH}(2.88) = 11.2$, $R_{\rm pVH}(2.71) = 9.5$ and $R_{\rm pVH}(2.58) = 8.2$, results comparable to those obtained previously by using rubbed PI layers\cite{Whitehead2000}. For non-aligned F6T2 (Figure~\ref{fig:FigurePLFilms}(e)), the PL vibronic peaks appear at $2.28\;$eV ($544\;$nm), $2.12\;$eV ($585\;$nm) and $1.95\;$eV ($636\;$nm) and $R_{\rm VH}$ increases from 1.16 to 3.50 (Figure~\ref{fig:FigurePLFilms}(g)), a ratio harder to calculate due to a weaker emission (in which the noise had to be removed using a fast Fourier transform algorithm) compared with F8BT and PFO.

\newpage

Al-Polymer-Al microcavities were fabricated using non-aligned F8BT, PFO, F6T2 and for each cavity, a corresponding Al-Aligned(SD1-Polymer)-Al microcavity was fabricated. Angle-resolved polarized reflectivity maps were recorded for each microcavity by varying the angle $\theta$ formed between the vector normal to the microcavity plane and the incident light direction. For the aligned cavities, the measurement was performed at an angle $\Phi$ (formed between TE polarization and the polymer chain direction) equal to $0^{\circ}$. All results were analysed using a Hopfield-Agranovich Hamiltonian\cite{Agranovich1957,Hopfield1958,Ciuti2005} including either one (PFO, F6T2) or two (F8BT) separate excitons with all fitting results displayed in Table \ref{tab:1}. For each microcavity, the experimental results are supported by transfer matrix reflectivity (TMR) calculations whose outputs are shown in Supporting Information. The reference results obtained for the non-aligned PFO and F8BT cavities are also shown in Supporting Information and agree with previous reports\cite{Tropf2017,LeRoux2018}. 

\begin{figure}[H]
\includegraphics[scale=0.75]{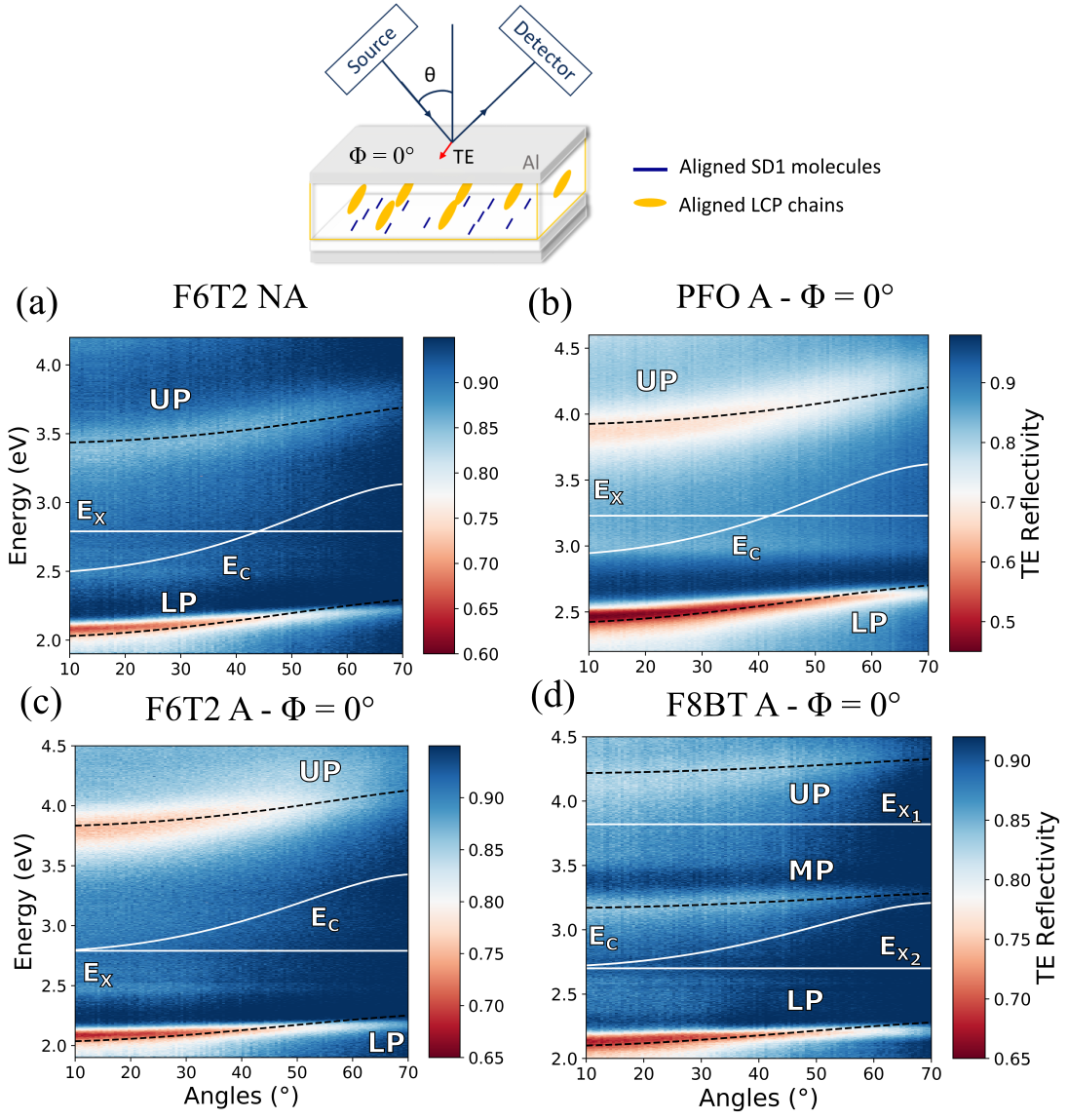}
\caption{\label{fig:FigureTECoupling} Experimental, angle-resolved, TE-polarized reflectivity maps for microcavities containing: (a) non-aligned F6T2, (b) aligned PFO, (c) aligned F6T2 and (d) aligned F8BT. For (b), (c) $\&$ (d), the measurements were performed at $\Phi = 0^{\circ}$ (see definition in the text) as depicted in the experiment schematic shown above the maps. Overlaid solid white lines are the exciton $E_{\rm X}$ and cavity $E_{\rm C}$ modes, black dashed lines are polaritons fitted from the analytical model.}
\end{figure}

\begin{figure}[H]
\includegraphics[scale=0.53]{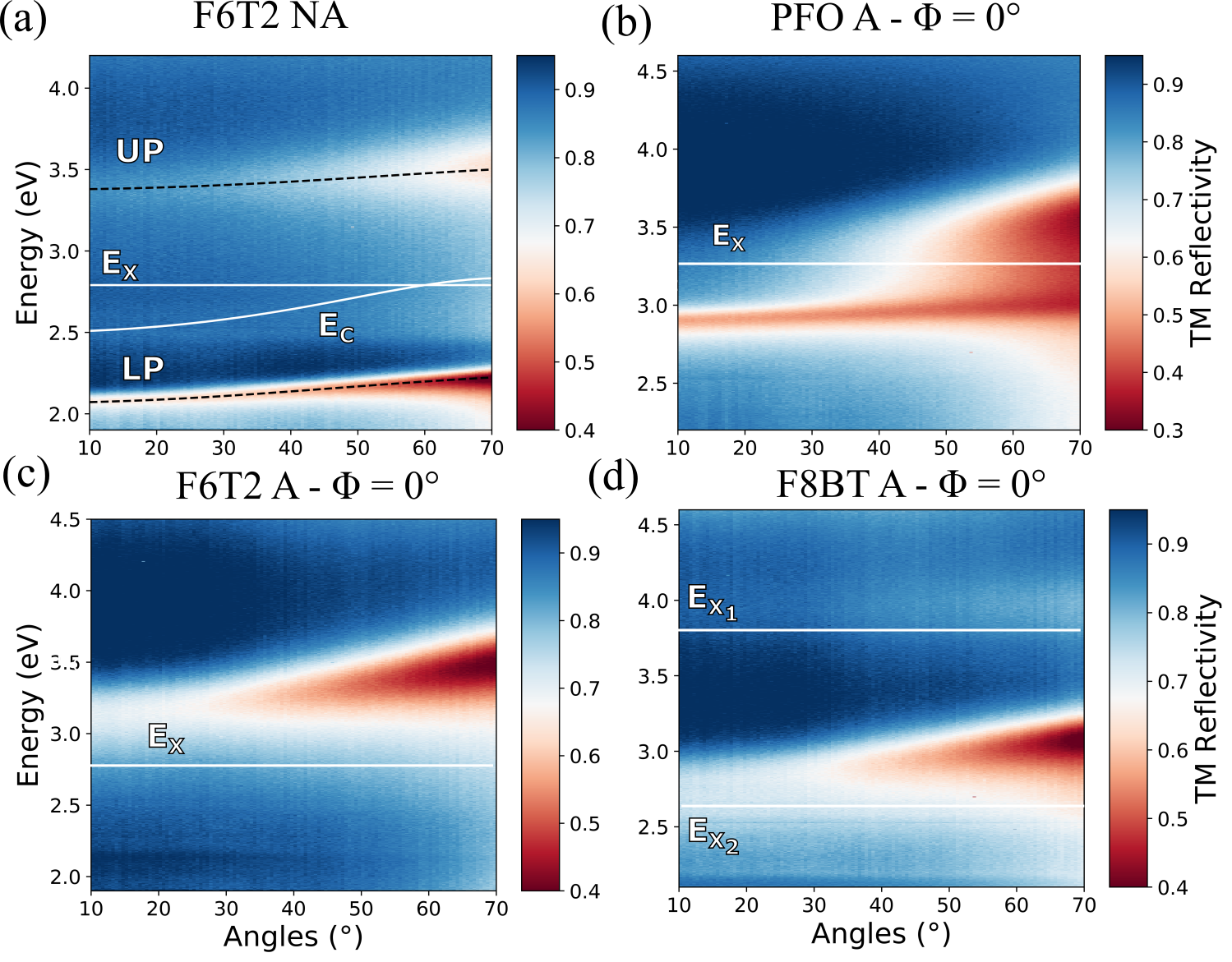}
\caption{\label{fig:FigureTMCoupling} Experimental, angle-resolved, TM-polarized reflectivity maps for microcavities containing (a) non-aligned F6T2 (b) aligned PFO (c) aligned F6T2 and (d) aligned F8BT. For (b), (c) $\&$ (d), the measurements were performed at $\Phi = 0^{\circ}$ (see definition in the text) and the overlaid solid white line is the exciton $E_{\rm X}$. In (a) in addition to the exciton $E_{\rm X}$, the second white line represents $E_{\rm C}$ the cavity mode and the black dashed lines the polaritons fitted from the analytical model.}
\end{figure}

Measured and fitted results for the non-aligned F6T2 cavity for TE polarization are shown in Figure~\ref{fig:FigureTECoupling} (a) where we observe the Lower (LP) and Upper (UP) Polaritons avoiding crossing of the exciton ($E_{\rm X_{\rm F6T2}} = 2.72\;$eV) by more than $1\;$eV, providing clear evidence of USC. Even though the layer is non-aligned, the largest Rabi splitting energy to date $\hbar\Omega_{0_{\rm TE}} = 1.38 \pm 0.01\;$eV is measured, corresponding to a coupling ratio $g = 49\%$. As expected, the results for TM polarization (Figure~\ref{fig:FigureTMCoupling} (a)) displays a flatter angular dispersion for the polaritons as the effective refractive index $n_{\rm eff_{\rm TM}}$ of the microcavity increases due to polarization-dependent penetration depth through the metallic mirrors\cite{Kena-Cohen2013,Economou1969,Litinskaya2012}. For this polarization, the in-plane/out-of-plane anisotropy of the active layer slightly reduces the Rabi energy down to $\hbar\Omega_{0_{\rm TM}} = 1.27 \pm 0.01\;$eV, as the contribution of the weaker out-of plane ($k_{\rm ex}$) component to the overall interaction increases along with $\theta$\cite{Kena-Cohen2013}.

Results for the aligned cavities of PFO, F6T2 and F8BT are shown respectively in Figure~\ref{fig:FigureTECoupling} (b), (c) and (d) for TE polarization at $\Phi = 0^{\circ}$. In each case, USC is clearly observed with fitting results revealing a systematic increase for $\hbar\Omega_{0_{\rm TE,\Phi = 0^{\circ}}}$ of $\sim 40\%$ compared to the non-aligned cavities. As the transition dipole moment $\boldsymbol{\mu}$ rotates together with the polymer chains in the $y$-direction, the relative increase in coupling strength can be calculated using the term $\boldsymbol{{\mu}.E}$ in Equation~\ref{eq:Scaling} to be: $\frac{\cos(0)}{\cos(\frac{\pi}{4})} = \sqrt{2}\sim 41 \%$ for TE polarization. Different factors can explain deviations to this theoretical value: (i) albeit thin, the SD1 layer can slightly decrease the overall coupling strength by diminishing the overlap between exciton and cavity modes, (ii) all the polymer chains are not perfectly aligned in the $y$ direction with remainding optical activity in the $x$ direction (see Figure \ref{fig:FigureIndices}) reducing the coupling strength, (iii) the reordering of the polymer can affect its microscopic properties. All the $\hbar\Omega_{0_{\rm TE,\Phi = 0^{\circ}}}$ values derived here, if not for the one we have just reported for non-aligned F6T2, exceed previous reports: the aligned F8BT cavity exhibits $\hbar\Omega_{01_{\rm TE,\Phi = 0^{\circ}}} = 1.18 \pm 0.01\;$eV and $\hbar\Omega_{02_{\rm TE,\Phi = 0^{\circ}}} = 1.25 \pm 0.01\;$eV, the aligned PFO cavity $\hbar\Omega_{0_{\rm TE,\Phi = 0^{\circ}}} = 1.47 \pm 0.01\;$eV and the aligned F6T2 cavity reaches a value of $\hbar\Omega_{0_{{\rm TE},\Phi = 0^{\circ}}} = 1.80 \pm 0.01\;$eV. To our knowledge, this last value is the first Rabi splitting energy to reach values comparable to photons in the visible spectrum ($\sim 689\;$nm) in a solid-state system, exceeding by more than 60$\%$ the previous record\cite{Liu2015, Liu2019} and corresponds to a slightly higher coupling ratio $g \sim 65 \%$ than the one obtained in Ref 25, fact which is remarkable since this result is obtained for an exciton lying $\sim 1.6\;$ eV higher in energy. 

\newpage

The results for TM polarization for the aligned cavities are shown in Figure~\ref{fig:FigureTMCoupling} (b), (c) and (d). In those measurements, the weak coupling regime is clearly observed for F6T2 (c) and F8BT (d) as the LP and UP are no longer visible and replaced by a single photonic mode. For PFO (b), the reflectivity observed does not show proper anti-crossing around the exciton ($E_{\rm X_{\rm PFO}} = 3.23\;$eV) with the PL measurements in the next section supporting the lack of evidence for SC. 

\begin{table}

\renewcommand{\arraystretch}{1.3}

\caption{\label{tab:1}Extracted and pre-set parameter values for microcavities containing non-aligned and aligned polymers, modelled using a Hopfield-Agranovich Hamiltonian\cite{Hopfield1958,Agranovich1957,Ciuti2005}. Values are shown for both TE- and TM-polarization, at $\Phi = 0^{\circ}$ (see definition in the text) for the aligned cavities. The initials A and NA respectively designate aligned and non-aligned polymer layers.}
\centering

\setlength\tabcolsep{3pt}

\begin{threeparttable}

\begin{tabular}{c|c|c|c|c|c|c}

Polymer: & PFO NA & PFO A & F8BT NA & F8BT A &	F6T2 NA & F6T2 A\\
$\hbar\omega_{1} ({\rm eV})$\tnote{1}$\;$ &	3.23 &	3.23 & 3.82 & 3.82 & 2.79 & 2.79\\
	
$\hbar\omega_{2}({\rm eV})$\tnote{2} $\;$&	- &	- & 2.70 & 2.70 & - & -\\

$\hbar\Omega_{01_{\rm TE}} ({\rm eV})$\tnote{3} $\;$& 1.02 $\pm$ 0.01 & 1.47 $\pm$ 0.01 &	0.84 $\pm$ 0.01 & 1.18 $\pm$ 0.01 & 1.38 $\pm$ 0.01 & 1.80 $\pm$ 0.01\\

$g_{01_{\rm TE}} \%$\tnote{4} $\;$& 32  & 46 &	22 & 31 & 49 & 65\\

$\hbar\Omega_{01_{\rm TM}} ({\rm eV})$\tnote{5} $\;$& 0.97 $\pm$ 0.01 & - & 0.80 $\pm$ 0.01 & - & 1.27 $\pm$ 0.01 & -\\

$g_{01_{\rm TM}} \%$\tnote{6} $\;$& 30  & - &	21 & - & 46 & -\\

$\hbar\Omega_{02_{\rm TE}} ({\rm eV})$\tnote{7}$\;$	& - & - & 0.84 $\pm$ 0.01 & 1.25 $\pm$ 0.01 & - & -\\

$g_{02_{\rm TE}} \%$\tnote{8} $\;$& -  & - & 31 & 46 & - & -\\

$\hbar\Omega_{02_{\rm TM}} ({\rm meV})$\tnote{9}$\;$& - & - & 0.80 $\pm$ 0.01 & - & - & -\\

$g_{02_{\rm TM}} \%$\tnote{10} $\;$& -  & - & 30 & - & - & -\\

$ n_{\rm eff_{\rm TE}}$\tnote{11} $\;$& 1.68 $\pm$ 0.01 & 1.70 $\pm$ 0.01 & 1.70 $\pm$ 0.01 & 1.87 $\pm$ 0.01	& 1.64 $\pm$ 0.01	& 1.71 $\pm$ 0.01\\

$ n_{\rm eff_{\rm TM}}$\tnote{12}$\;$ & 2.33 $\pm$ 0.01 & - & 2.37 $\pm$ 0.01 & - & 2.12 $\pm$ 0.01	& -\\

$ E_{\rm c_{\rm TE}} (0) ({\rm eV})$\tnote{13} $\;$&	3.13 $\pm$ 0.02	& 2.92 $\pm$ 0.02 & 2.54 $\pm$ 0.02 &	2.71 $\pm$ 0.02	& 2.48 $\pm$ 0.02 & 2.78 $\pm$ 0.02\\ 

$ E_{\rm c_{\rm TM}} (0)({\rm eV})$\tnote{14}	$\;$& 3.15 $\pm$ 0.02	& - & 2.59 $\pm$ 0.02 & - & 2.50 $\pm$ 0.02 & -\\

\end{tabular}

\begin{tablenotes}

\small

\item[1] Exciton oscillator 1 transition energy.
\item[2] Exciton oscillator 2 transition energy.
\item[3] TE-polarized Rabi energy associated with exciton 1 for $\omega_{\rm cav_{\rm TE}}=\omega_{1}$ (see definition in the text).
\item[4] TE-polarized  normalized coupling ratio energy associated with exciton 1 (see definition in the text).
\item[5] TM-polarized Rabi energy associated with exciton 1 for $\omega_{\rm cav_{\rm TM}}=\omega_{1}$ (see definition in the text).
\item[6] TM-polarized  normalized coupling ratio energy associated with exciton 1 (see definition in the text).
\item[7] TE-polarized Rabi energy associated with exciton 2 for $\omega_{\rm cav_{\rm TE}}=\omega_{1}$ (see definition in the text).
\item[8] TE-polarized  normalized coupling ratio energy associated with exciton 2 (see definition in the text).
\item[9] TM-polarized Rabi energy associated with exciton 2 for $\omega_{\rm cav_{\rm TM}}=\omega_{1}$ (see definition in the text).
\item[10] TM-polarized  normalized coupling ratio energy associated with exciton 2 (see definition in the text).
\item[11] Effective refractive index for TE polarization.
\item[12] Effective refractive index for TM polarization.
\item[13] TE-polarized energy of the bare cavity mode at normal incidence.
\item[14] TM-polarized energy of the bare cavity mode at normal incidence.

\end{tablenotes}

\end{threeparttable}
\end{table}

\newpage

Angle-resolved PL was recorded for each microcavity. The excitation laser used for the thin film PL measurement was focused onto the sample at an incidence of 75$^{\circ}$, the excitation geometry between films and microcavities experiments being identical. For each microcavity, the average power was kept low ($\leq 10 \mu$W) and the excitation energy was chosen to optically pump one of the intense absorptions arising from the optical transitions of the underlying polymer. For the aligned cavities, the measurement was performed at $\Phi = 0^{\circ}$. 

\begin{figure}[H]
\begin{flushleft}
\includegraphics[scale=0.23]{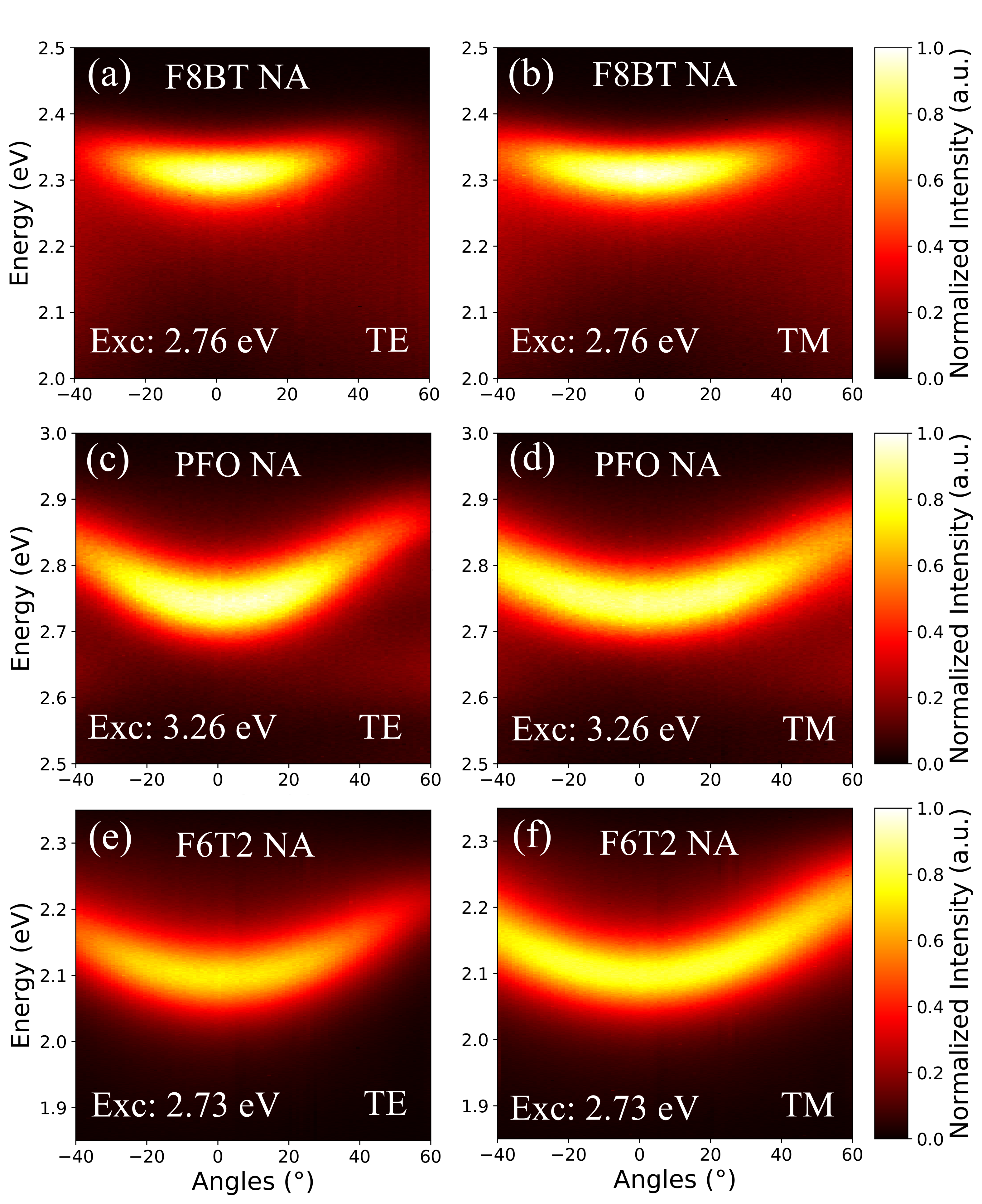}
\end{flushleft}
\caption{\label{fig:FigureAPLNA} Angle-resolved PL spectral intensity maps for microcavities containing non-aligned F8BT ((a) $\&$ (b)), PFO ((c) $\&$ (d)) and F6T2 ((e) $\&$ (f)), with TE ((a), (c), (e)) and TM ((b), (d), (f)) polarized spectra plotted separately. The excitation energy is overlaid in white for each measurement. Each polarization pair ((a) $\&$  (b)), ((c) $\&$ (d)), ((e) $\&$  (f)) is normalized by the same value to allow for comparison.}
\end{figure}

PL intensity maps plotted by energy vs emission angle for both TE- ((a), (c) and (e))) and TM-polarized ((b), (d) and (f)) emissions from the non-aligned F8BT ((a) and (b)), PFO ((c) and (d)) and F6T2 ((e) and (f)) microcavities are shown in Figure~\ref{fig:FigureAPLNA}. In each case the emission is dominated by a narrow single peak originating from the LP which is relatively insensitive to angular-dispersion: a recognizable feature of USC\cite{Kena-Cohen2013,Gambino2014,Mazzeo2014} (values including peak positions and full width at half maximum (FWHM) at normal incidence as well as the angular dispersion from 0 to $60^{\circ}$ are shown in Table \ref{tabPL}). The results for PFO agree with a previous report\cite{LeRoux2018} and the variety of polymers used here allows for emissions across the visible spectrum (in the blue at $\sim 452\;$ nm for PFO, green at $\sim 537\;$nm for F8BT and yellow/orange at $\sim 588\;$nm for F6T2). 

\begin{figure}[H]
\begin{flushleft}
\includegraphics[scale=0.18]{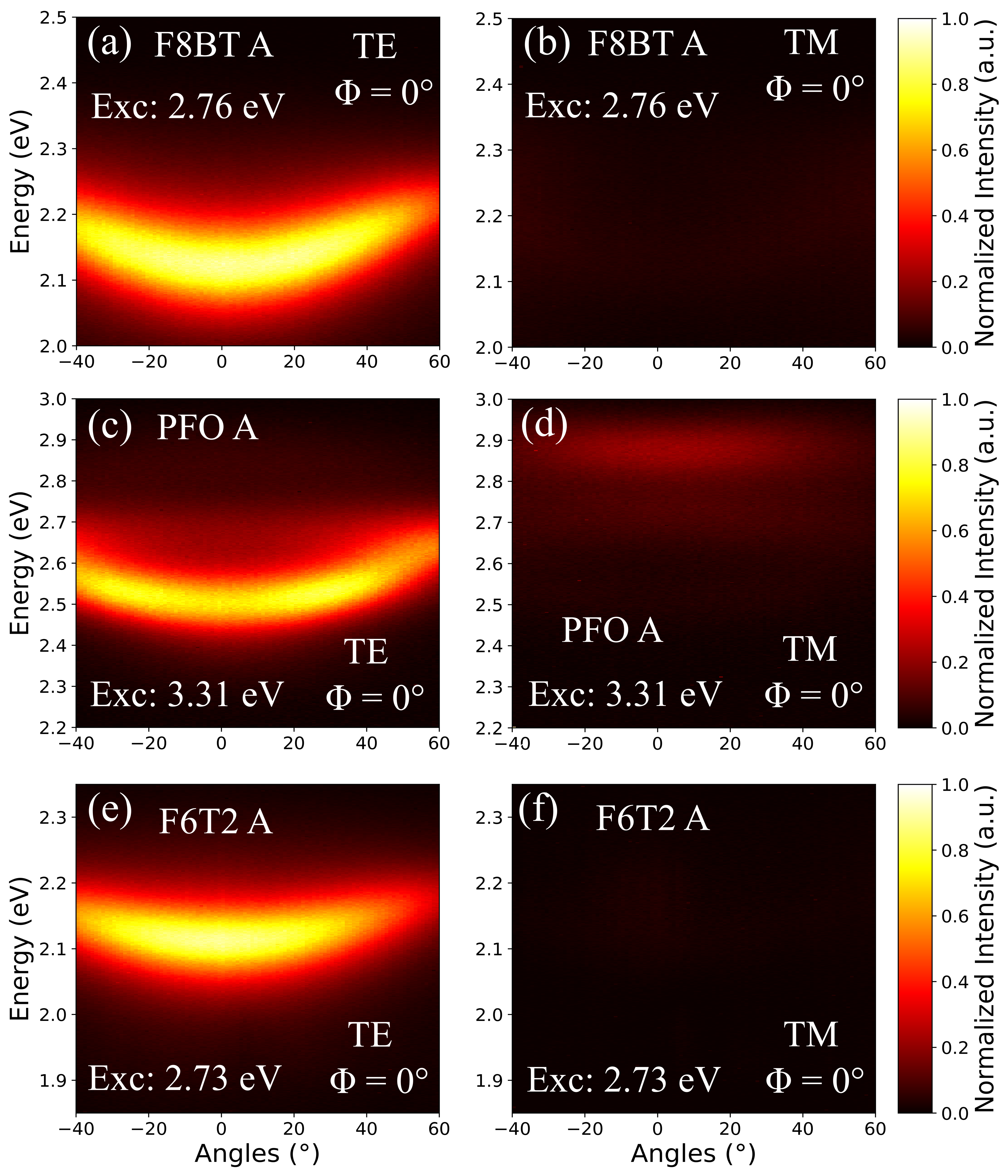}
\end{flushleft}
\caption{\label{fig:FigureAPLA} Angle-resolved PL spectral intensity maps for microcavities containing aligned F8BT ((a) $\&$ (b)), PFO ((c) $\&$ (d)) and F6T2 ((e) $\&$ (f)), with TE ((a), (c), (e)) and TM ((b), (d), (f)) polarized spectra plotted separately. The measurements were performed at $\Phi = 0^{\circ}$ (see definition in the main text) . Each polarization pair ((a) $\&$ (b)), ((c) $\&$ (d)), ((e) $\&$ (f)) is normalized by the same value to allow for comparison.}
\end{figure}

Corresponding PL intensity maps for the aligned cavities are shown in Figure~\ref{fig:FigureAPLA}. TE-polarized emission ((a), (c) and (f)) resembles the one observed from non-aligned microcavities with a single peak emitted from the LP. It however differs in energy for F8BT and PFO between non-aligned and aligned cavities: the peak emission is recorded at $2.13\;$eV at normal incidence for aligned F8BT compared to $2.31\;$eV when non-aligned, and at $2.59\;$eV for aligned PFO compared to $2.74\;$eV when non-aligned. These redshifts are not the result of different thicknesses between the cavities as TMR calculations show that for each pair, the polymer layer thicknesses are comparable (the aligned F8BT is $110\;$nm-thick compared to $118\;$nm when aligned, the aligned PFO 96 nm-thick compared to 97 nm when aligned) but are direct evidence of the increased interaction strength which repels the UP and LP to respectively higher and lower energies. The emission for aligned and non-aligned F6T2 is closer in energy at normal incidence ($2.11\;$eV aligned compared with $2.10\;$eV non-aligned) and is this time the result of a much larger thickness of the non-aligned F6T2 layer (the aligned F6T2 is $94\;$nm-thick compared to $123\;$nm non-aligned), resulting in lower energy cavity modes and LP (as can be observed in Figure~\ref{fig:FigureTECoupling}) which makes up for the difference in interaction strength between the two microcavities.

TM-polarized measurements ((b), (d) and (f)) show no distinguishable emission for F8BT (b) and F6T2 (g) as the only photonic mode in both those cavities lies too high in energy (at $\sim 3\;$eV for F6T2 and $\sim 2.9\;$eV for F8BT) to allow any emission from the underlying polymer (Figure \ref{fig:FigurePLFilms} (b) and (f) shows the emission located at energies lower than $2.6\;$eV for both F6T2 and F8BT). For the microcavity containing aligned PFO (d), a  5-times weaker, broad, angle-independent TM-polarized emission was detected (f), with two maxima at 2.88 and $2.71\;$eV coinciding with the ones from the bare film (Figure \ref{fig:FigurePLFilms}(d)). This broad, structured, angle-insensitive dispersion confirms that the microcavity no longer operates under USC for this polarization. The emission itself is only allowed through the photonic mode at $\sim 2.88\;$eV (see \ref{fig:FigurePLFilms}(b)) and even though this photonic mode overlaps with the most intense part of the bare film's PL, the resulting signal is much weaker in TM than TE polarization as the emission in the direction perpendicular to the chain alignment is intrinsically much weaker. 
 
\newpage

\newpage

\begin{table}
\caption{\label{tabPL} Peak positions and FWHMs at normal incidence for the TE- and TM-polarized emission
s displayed in Figure \ref{fig:FigureAPLNA} and \ref{fig:FigureAPLA}. The angular dispersion of the emission for both polarizations from 0 to $60^{\circ}$ is also reported. The initials A and NA respectively designate aligned and non-aligned polymer layers.}
\renewcommand{\arraystretch}{1.3}
\centering
\setlength\tabcolsep{3pt}

\begin{tabular}{c|c|c|c|c|c|c}
Polymer: & PFO NA & PFO A & F8BT NA & F8BT A &	F6T2 NA & F6T2 A\\

Peak position (eV)&	2.74 &	2.59 & 2.31 & 2.13 & 2.10 & 2.11\\

${\rm FWHM}_{\rm TE}$ (meV)&	99 $\pm$ 1 & 140 $\pm$ 5 & 89 $\pm$ 5 & 127 $\pm$ 1 & 109 $\pm$ 1 & 109 $\pm$ 1\\

${\rm FWHM}_{\rm TM}$ (meV)&	99 $\pm$ 1 & - & 82 $\pm$ 5 & - & 109 $\pm$ 1 & -\\
	
TE dispersion (meV) & 130 & 50 & 30 & 80 & 110 & 60\\

TM dispersion (meV)& 110 & - & 20 & - & 110 & -\\

\end{tabular}

\end{table}

Several opportunities unique to our photoalignment technique for exciton-polaritons exist: the possibility to align SD1 at mesoscopic scales\cite{Ma2015, Ma2017, Zhang2019} allows local alignment of the polymer chains, which could lead to the fabrication of the missing NOT gate in optical logic thanks to spontaneous splittings between TE and TM polarizations\cite{Sanvitto2016,Espinosa-Ortega2013,Solnyshkov2015}, rapid switching of the coupling strength using LCPPs and LCs in microcavities could also be an opportunity to perform direct extraction of the USC ground state virtual contents\cite{Ciuti2005, FriskKockum2019, Forn-Diaz2019} and the fabrication of complex energy landscapes, especially when combined with the tunability of the molecular structure (for example by generating segments of $\beta$-phase in PFO\cite{Perevedentsev2015}) could help address challenges in quantum simulation\cite{Scafirimuto2018}. On a more practical standpoint, we expect that a wide breadth of polarization sensitive devices and phenomena such as Bose-Einstein condensation and exciton-polariton lasing (which has recently been demonstrated using pentafluorene \cite{Rajendran2019}) will take advantage of the polarization-dependent coupling (the spontaneous polarization observed during lasing and condensation could for instance be controlled by molecular alignment).

\section{Conclusions}

We have fabricated organic microcavities containing LCCPS (F8BT, PFO, F6T2) aligned using a thin photoalignment layer (SD1). The USC regime was first observed for the non-aligned microcavities with a value of $\hbar\Omega_{\rm R_{F6T2}} = 1.34$ eV exceeding previous reports. The alignment then allowed for a systematic increase of the coupling strength in the direction of the alignment with giant values of $\hbar\Omega_{\rm R}$ culminating at 1.80 eV for F6T2, a value comparable to photon energies in the visible spectrum, also corresponding to the highest coupling ratio $g$ = 65$\%$ to date in the solid-state\cite{FriskKockum2019,Forn-Diaz2019}. Angle-resolved PL for the TE polarization parallel to the alignment direction revealed red-shifted LP emissions compared to the non-aligned cavities, a signature of the increased interaction strength. In this geometry, the absence of polaritons in TM-polarized reflectivity and weak or no PL also demonstrated that the coupling strength was polarization dependent. By using three different polymers, we demonstrated that the alignment can be generalized to other LCPPs and that a real opportunity to reach coupling ratios close to $90\%$ exists if the alignment can be applied at lower energies. Aligning LCCPS in microcavities at microscopic scales using SD1 also offers further possibilities for the realization of polaritonic devices and rich energy landscapes. 

\section{Methods}

\subsection{Materials}

The three polymers used in this study were supplied by Cambridge Display Technology (F8BT), Sumitomo Chemical (PFO and F6T2) and used as received. Their peak molecular weights were: $M_{p_{\rm PFO}}$ = $50\times 10^{3}$ g.mol$^{-1}$, $M_{p_{\rm F8BT}}$ = $77\times 10^{3}$ g.mol$^{-1}$, $M_{p_{\rm F6T2}}$ = $80\times 10^{3}$ g.mol$^{-1}$. The azo-dye photo-alignement layer SD1 was supplied by Dai-Nippon Ink and Chemicals, Japan. Anhydrous toluene (99.8$\%$), anhydrous chloroform ($\geq$99$\%$) and anhydrous 2-methoxyethanol($\geq$99.8$\%$) were purchased from Sigma-Aldrich. Solvents were used as received. For the mirror fabrication, Aluminium pellets (99.999$\%$) were purchased from Kurt J. Lesker.

\subsection{Film Fabrication}

The bare films (used for ellipsometry and PL) of SD1, PFO, F8BT, F6T2 were spincoated from solutions in 2-methoxyethanol (SD1 at $1$ mg.mL$^{-1}$), toluene (PFO at $18\;$mg.mL$^{-1}$ and F8BT at $18\;$mg.mL$^{-1}$) and chloroform (F6T2 at $13\;$mg.mL$^{-1}$). All solutions  were prepared in an inert environment, left to stir overnight at a temperature of $55^{\circ}$C except for F6T2 in chloroform which was left stirring at room temperature. All solutions were then filtered using a $0.45 \;\mu$m PTFE filter. All samples were spincoated on top of fused silica substrates. The non-aligned polymer films were spun for $1\;$min at a speed of $2000\;$rpm with an acceleration of $1200\;$rpm.s$^{-1}$. For the aligned polymer films, the SD1 layer was first spincoated by spinning $5\;$s at $500\;$rpm (acceleration $500\;$rpm.s$^{-1}$) and $25\;$s at $2000\;$rpm (acceleration $1200\;$rpm.s$^{-1}$). The film was then annealed for 6 minutes at a temperature of $150^{\circ}$C to drive any traces of solvent away. The alignment of the SD1 layer was performed in air by exposing the sample to $5\;$mW of polarized UV light (emitted by a M365LP1 LED in front of a broadband WP25M-UB polarizer from Thorlabs) for 10 minutes; The SD1 chains aligning perpendicular to the direction of the polarized light\cite{Li2006, Ma2015}. The samples were then put back in an inert environment where the polymer layer was spincoated by spinning for $1\;$min at a speed of $2000\;$rpm with an acceleration of $1500\;$rpm.s$^{-1}$. Each polymer was then thermally annealed into their respective nematic phases (160$^{\circ}$C for PFO, 220$^{\circ}$C for F6T2, 250$^{\circ}$C for F8BT) using a Linkam THMS600 heating stage with a heating rate of 30$^{\circ}$C.min$^{-1}$. This temperature was held for 10 minutes and subsequent quenching to room temperature was realized by quickly placing the sample on the metallic floor of the glovebox while applying a gentle flow of nitrogen.

\subsection{Microcavity Fabrication}

The aluminium mirrors were evaporated at a rate of $10\;$A.s$^{-1}$ at a pressure of $10^{-9}\;$mbar. For the non-aligned cavities, the spincoating conditions used on top of the bottom mirror were identical to the ones described for the bare films. For the aligned cavities, the SD1 layer was spincoated on top of the bottom mirror. The structures fabricated had low Q factors $\sim 25$, characteristic of metallic microcavities using Aluminium. The concentration of the SD1 solution in 2-methoxyethanol was increased to $3$ mg.mL$^{-1}$ as spincoating SD1 on a metallic surface results in lower thicknesses than on fused silica. A slightly thicker layer also acts as a protection layer to prevent the aluminium from reacting with the polymer upon annealing at high temperatures (using $1\;$mg.mL$^{-1}$ SD1 solution resulted in samples unfit for measurement). The sample was then annealed 6 minutes at a temperature of $150^{\circ}$C to drive any traces of solvent away and the rest of the alignment procedure was similar to the one used for the bare films. For the spincoating of the polymer layers, the solution concentrations were in some cases adjusted so as to adjust the thicknesses and therefore the cavity mode energy. The concentrations used in the non-aligned microcavities were: PFO at $19\;$mg.mL$^{-1}$ , F8BT at $20\;$mg.mL$^{-1}$ and F6T2 at $15\;$mg.mL$^{-1}$. The concentrations for the aligned microcavities were: PFO at $18\;$mg.mL$^{-1}$, F8BT at $18\;$mg.mL$^{-1}$, F6T2 at $13\;$mg.mL$^{-1}$.

\subsection{Optical Characterization}

The optical constants for the non-aligned and aligned films of PFO, F6T2 and F8BT were extracted using a J.A. Woollam ESM-300 ellipsometer. For each sample, 8 reflection-geometry measurements were performed with light incident from 45$^{\circ}$ to 61$^{\circ}$ (angles of incidence are quoted relative to the plane normal) together with a normal incidence (0$^{\circ}$) transmission measurement, for the aligned films the measurement was performed at $\Phi = 0^{\circ}$. The reflectivity maps obtained in Figure \ref{fig:FigureIndices} were obtained using a home-built white light reflectivity setup. The microcavities were placed at the center of a stage with two independent rotating arms. A deuterium-halogen light source (DH-2000-DUV from Ocean Optics) was coupled into a fiber whose output was collimated onto a broadband polarizer (WP25M-UB from Thorlabs) and onto the sample (final spot size $1\;$mm). The reflected light was then coupled into a second fiber placed onto the second arm and analyzed using a spectrometer (HRS 500, 150 g/mm grating with blazing wavelength at 300 nm) and CCD (Pylon-2KB CCD from Princeton Instruments). In all cases, a neat aluminium mirror with known reflectivity was used as reference. The acquisition angle was varied by $0.5^{\circ}$ steps from 10 to 70$^{\circ}$.

\subsection{Photoluminescence}

All the time-integrated PL measurements were performed on the same home-built setup. The fiber coupled to the white light source was disconnected and the first arm was positioned at 75$^{\circ}$ relative to the sample plane normal. The pulsed laser beam from a supercontinuum white light laser (SuperK Extreme with its UV spectral extension unit Extend-UV, NKT Photonics) was used as excitation source and focused onto the sample (spot size $<1$ mm). The excitation energy was tuned according to the optical transitions, the incident power was kept low ($\leq 10\mu W$) with pulse widths of 20 to 30 ps and a repetition rate of $77.87\;$MHz. The broadband polarizer was placed on the collection arm, at a distance of $10\;$ cm from the sample. Two nearly closed irises ($\sim 1$cm) at a distance of $5\;$cm from each other were then placed before the coupling lens of the collection fiber in order to ensure that the collected light was emitted at the desired angle in the horizontal plane. The light was then analysed using the spectrometer and CCD described before using this time a 300 g/mm grating  blazed at a wavelength of $500\;$nm. For the polymer films, the light was collected at normal incidence. For the microcavities the acquisition angle was varied by $1^{\circ}$ steps from -40 to 60$^{\circ}$.

\subsection{Data Analysis}

The minima of the reflectivity maps shown in Figures \ref{fig:FigureTECoupling} and \ref{fig:FigureTMCoupling} were analyzed using a least-square fitting algorithm for the eigenvalue problem:

\begin{equation}\label{eq:8}
H_{q} \bold{v_{i,q}}= \omega_{i,q} \bold{v_{i,q}},
\end{equation}

where $H_q$ is an extension of the Agranovich/Hopfield Hamiltonian containing either one\cite{Hopfield1958, Agranovich1957, Ciuti2005} (PFO, F6T2) or two excitonic resonances\cite{LeRoux2018} (F8BT):

\begin{equation}\label{eq:9}
H_{q}=
\begin{bmatrix}
\omega_{{\rm cav},q}+2D_{q} & -i\frac{\Omega_{1,q}}{2} & -i\frac{\Omega_{2,q}}{2} & -2D_{q} & -i\frac{\Omega_{1,q}}{2} & -i\frac{\Omega_{2,q}}{2}\\ 
i\frac{\Omega_{1,q}}{2} & \omega_{1} & 0 & -i\frac{\Omega_{1,q}}{2} & 0 & 0\\
i\frac{\Omega_{2,q}}{2} & 0 & \omega_{2} & -i\frac{\Omega_{2,q}}{2} & 0 & 0\\ 
2D_{q} & -i\frac{\Omega_{1,q}}{2} & -i\frac{\Omega_{2,q}}{2} & -\omega_{{\rm cav},q} - 2D_{q} & -i\frac{\Omega_{1,q}}{2} & -i\frac{\Omega_{2,q}}{2}\\
-i\frac{\Omega_{1,q}}{2} & 0 & 0 & i\frac{\Omega_{1,q}}{2} & -\omega_{1} & 0\\ 
-i\frac{\Omega_{2,q}}{2} & 0 & 0 & i\frac{\Omega_{2,q}}{2} & 0 & -\omega_{2}
\end{bmatrix}   ,
\end{equation} 

In the case of a single exciton oscillator, $H_q$ reduces to the usual 4 x 4 Hopfield-like USC matrix\cite{Kena-Cohen2013}. In Eq.~(\ref{eq:8}),~(\ref{eq:9}), q is the in-plane wave vector, $\omega_{{\rm cav}_q}$ the cavity mode energy, $\omega_j$ the frequency for the $j$-excitons,  $\Omega_{j,q}$ is the associated Rabi frequency, and for a given angle $\theta$: $\Omega_{j,q}= \Omega_{j}(\theta) = \Omega_{0j}\sqrt{\frac{\omega_j}{\omega_{{\rm cav}}(\theta)}}$ where $\Omega_{0j}$ is the Rabi frequency on resonance for the $j$-excitons. It was shown that in metal-organic semiconductor-metal cavities $\omega_{{\rm cav}}(\theta)$ can be approximated by\cite{Kena-Cohen2013}:
 
\begin{equation}\label{eq:5}
\omega_{{\rm cav}_{{\rm (TE,TM)},q}} = \omega_{{\rm cav}_{{\rm (TE,TM)}}}(\theta)= 
\omega_{{\rm cav}}(0)\left(1-\frac{\sin^{2}(\theta)}{n^{2}_{{\rm eff}_{{\rm (TE,TM)}}}}\right)^{-\frac{1}{2}}
\end{equation}
where $n_{\rm eff_{\rm TE,TM}}$ is polarization dependent. Finally, $D_q = \sum _{j} \frac{\Omega_{j,q}^2}{4\omega_{j}}$ is the contribution of the squared magnetic vector potential.

In order to diagonalize \textit{H}, the polariton annihilation operators $p_{i,q} = w_{i,q}a_{q}+\sum_{j}x_{i,j,q}b_{j,q}+y_{i,q}a^{\dagger}_{-q}+\sum_{j}z_{i,j,q}b^{\dagger}_{j,-q}$ for $i\epsilon\left \{\rm LP,MP,UP\right\}$ are introduced, where $a_q$  and $a^{\dagger}_{q}$ respectively annihilate and create a photon at frequency $\omega_{{\rm cav}_q}$, $b_j$  and $b^{\dagger}_j$ respectively annihilate and create a $j$-exciton at frequency $\omega_j$. The terms $w, x, y$ and $z$ label, respectively, the photon, exciton, anomalous photon and anomalous exciton Hopfield coefficients. The eigenvalues of $H_q$ were fitted to the experimental results for each cavity, for both TE- and TM-polarization, using the $R$-minima in the 10 - 70$^{\circ}$ range. 

In order to minimize the number of fitting parameters and obtain meaningful results, only $\omega_{{\rm cav}_{\rm TE,TM}}(0)$, $n_{\rm eff_{\rm TE,TM}}$  $\&$  $\Omega_{01_{\rm TE,TM}}$ were allowed to vary in fittings of the PFO and F6T2 cavities. Similarly, only $\omega_{{\rm cav}_{\rm TE,TM}}(0)$, $n_{\rm eff_{\rm TE,TM}}$, $\Omega_{01_{\rm TE,TM}}$ $\&$  $\Omega_{02_{\rm TE,TM}}$ were allowed to vary in the fitting of the F8BT cavities. For each exciton, the value of $\hbar\omega_{j}$ was set to be at the energy that corresponds to the mid-point of the integral oscillator strength for the corresponding optical transition using $\int_{E_{\rm min}}^{\hbar\omega_{j}}\epsilon(\omega)d\omega=\frac{1}{2}\int_{E_{\rm min}}^{E_{\rm max}}\epsilon(\omega)d\omega$, where $\epsilon(\omega)$ is the extinction coefficient for $X_{j}$ in the $E_{\rm min}$ to $E_{\rm max}$ energy range. 

%\subsection{Authors Contributions}
%
%D.D.C.B. proposed using SD1 as a photoalignment layer for LCCPs. D.D.C.B. and F.L.R discussed their use in optical microcavities. F.L.R. carried out the fabrication, characterization experiments and the analysis of the data. R.A.T. assisted with the optical spectroscopy measurements. D.D.C.B., R.A.T and F.L.R wrote the manuscript.

\begin{acknowledgement}

The authors thank Prof. Moritz Riede for access to his facilities and Dr Richard Hamilton for fruitful discussions. They also acknowledge funding from the University of Oxford, from the UK Engineering and Physical Sciences Research Council and the Jiangsu Industrial Technology Research Institute. F.L.R. further thanks Wolfson College and Dr Simon Harrison for the award of a Wolfson Harrison UK Research Council Physics Scholarship.

\end{acknowledgement}

\bibliography{manuscriptACSbiblio}

\providecommand{\latin}[1]{#1}
\makeatletter
\providecommand{\doi}
  {\begingroup\let\do\@makeother\dospecials
  \catcode`\{=1 \catcode`\}=2 \doi@aux}
\providecommand{\doi@aux}[1]{\endgroup\texttt{#1}}
\makeatother
\providecommand*\mcitethebibliography{\thebibliography}
\csname @ifundefined\endcsname{endmcitethebibliography}
  {\let\endmcitethebibliography\endthebibliography}{}
\begin{mcitethebibliography}{66}
\providecommand*\natexlab[1]{#1}
\providecommand*\mciteSetBstSublistMode[1]{}
\providecommand*\mciteSetBstMaxWidthForm[2]{}
\providecommand*\mciteBstWouldAddEndPuncttrue
  {\def\EndOfBibitem{\unskip.}}
\providecommand*\mciteBstWouldAddEndPunctfalse
  {\let\EndOfBibitem\relax}
\providecommand*\mciteSetBstMidEndSepPunct[3]{}
\providecommand*\mciteSetBstSublistLabelBeginEnd[3]{}
\providecommand*\EndOfBibitem{}
\mciteSetBstSublistMode{f}
\mciteSetBstMaxWidthForm{subitem}{(\alph{mcitesubitemcount})}
\mciteSetBstSublistLabelBeginEnd
  {\mcitemaxwidthsubitemform\space}
  {\relax}
  {\relax}

\bibitem[Weisbuch \latin{et~al.}(1992)Weisbuch, Nishioka, Ishikawa, and
  Arakawa]{Weisbuch1992}
Weisbuch,~C.; Nishioka,~M.; Ishikawa,~A.; Arakawa,~Y. {Observation of the
  coupled exciton-photon mode splitting in a semiconductor quantum
  microcavity}. \emph{Physical Review Letters} \textbf{1992}, \emph{69},
  3314--3317\relax
\mciteBstWouldAddEndPuncttrue
\mciteSetBstMidEndSepPunct{\mcitedefaultmidpunct}
{\mcitedefaultendpunct}{\mcitedefaultseppunct}\relax
\EndOfBibitem
\bibitem[Marks \latin{et~al.}(1994)Marks, Halls, Bradley, Friend, and
  Holmes]{Marks1994}
Marks,~R.~N.; Halls,~J. J.~M.; Bradley,~D. D.~C.; Friend,~R.~H.; Holmes,~A.~B.
  {The photovoltaic response in poly(p-phenylene vinylene) thin-film devices}.
  \emph{Journal of Physics: Condensed Matter} \textbf{1994}, \emph{6},
  1379--1394\relax
\mciteBstWouldAddEndPuncttrue
\mciteSetBstMidEndSepPunct{\mcitedefaultmidpunct}
{\mcitedefaultendpunct}{\mcitedefaultseppunct}\relax
\EndOfBibitem
\bibitem[Alvarado \latin{et~al.}(1998)Alvarado, Seidler, Lidzey, and
  Bradley]{Alvarado1998}
Alvarado,~S.; Seidler,~P.; Lidzey,~D.~G.; Bradley,~D. D.~C. {Direct
  Determination of the Exciton Binding Energy of Conjugated Polymers Using a
  Scanning Tunneling Microscope}. \emph{Physical Review Letters} \textbf{1998},
  \emph{81}, 1082--1085\relax
\mciteBstWouldAddEndPuncttrue
\mciteSetBstMidEndSepPunct{\mcitedefaultmidpunct}
{\mcitedefaultendpunct}{\mcitedefaultseppunct}\relax
\EndOfBibitem
\bibitem[Plumhof \latin{et~al.}(2014)Plumhof, St{\"{o}}ferle, Mai, Scherf, and
  Mahrt]{Plumhof2014}
Plumhof,~J.~D.; St{\"{o}}ferle,~T.; Mai,~L.; Scherf,~U.; Mahrt,~R.~F.
  {Room-temperature Bose–Einstein condensation of cavity exciton–polaritons
  in a polymer}. \emph{Nature Materials} \textbf{2014}, \emph{13},
  247--252\relax
\mciteBstWouldAddEndPuncttrue
\mciteSetBstMidEndSepPunct{\mcitedefaultmidpunct}
{\mcitedefaultendpunct}{\mcitedefaultseppunct}\relax
\EndOfBibitem
\bibitem[Daskalakis \latin{et~al.}(2014)Daskalakis, Maier, Murray, and
  K{\'{e}}na-Cohen]{Daskalakis2014}
Daskalakis,~K.~S.; Maier,~S.~A.; Murray,~R.; K{\'{e}}na-Cohen,~S. {Nonlinear
  interactions in an organic polariton condensate}. \emph{Nature Materials}
  \textbf{2014}, \emph{13}, 271--278\relax
\mciteBstWouldAddEndPuncttrue
\mciteSetBstMidEndSepPunct{\mcitedefaultmidpunct}
{\mcitedefaultendpunct}{\mcitedefaultseppunct}\relax
\EndOfBibitem
\bibitem[Lerario \latin{et~al.}(2017)Lerario, Ballarini, Fieramosca, Cannavale,
  Genco, Mangione, Gambino, Dominici, {De Giorgi}, Gigli, and
  Sanvitto]{Lerario2017}
Lerario,~G.; Ballarini,~D.; Fieramosca,~A.; Cannavale,~A.; Genco,~A.;
  Mangione,~F.; Gambino,~S.; Dominici,~L.; {De Giorgi},~M.; Gigli,~G.;
  Sanvitto,~D. {High-speed flow of interacting organic polaritons}.
  \emph{Light: Science {\&} Applications} \textbf{2017}, \emph{6},
  e16212--e16212\relax
\mciteBstWouldAddEndPuncttrue
\mciteSetBstMidEndSepPunct{\mcitedefaultmidpunct}
{\mcitedefaultendpunct}{\mcitedefaultseppunct}\relax
\EndOfBibitem
\bibitem[Zasedatelev \latin{et~al.}(2019)Zasedatelev, Baranikov, Urbonas,
  Scafirimuto, Scherf, St{\"{o}}ferle, Mahrt, and Lagoudakis]{Zasedatelev2019}
Zasedatelev,~A.~V.; Baranikov,~A.~V.; Urbonas,~D.; Scafirimuto,~F.; Scherf,~U.;
  St{\"{o}}ferle,~T.; Mahrt,~R.~F.; Lagoudakis,~P.~G. {A room-temperature
  organic polariton transistor}. \emph{Nature Photonics} \textbf{2019},
  \emph{13}, 378--383\relax
\mciteBstWouldAddEndPuncttrue
\mciteSetBstMidEndSepPunct{\mcitedefaultmidpunct}
{\mcitedefaultendpunct}{\mcitedefaultseppunct}\relax
\EndOfBibitem
\bibitem[Lidzey \latin{et~al.}(1998)Lidzey, Bradley, Skolnick, Virgili, Walker,
  and Whittaker]{Lidzey1998}
Lidzey,~D.~G.; Bradley,~D. D.~C.; Skolnick,~M.~S.; Virgili,~T.; Walker,~S.;
  Whittaker,~D.~M. {Strong exciton–photon coupling in an organic
  semiconductor microcavity}. \emph{Nature} \textbf{1998}, \emph{395},
  53--55\relax
\mciteBstWouldAddEndPuncttrue
\mciteSetBstMidEndSepPunct{\mcitedefaultmidpunct}
{\mcitedefaultendpunct}{\mcitedefaultseppunct}\relax
\EndOfBibitem
\bibitem[Schwartz \latin{et~al.}(2011)Schwartz, Hutchison, Genet, and
  Ebbesen]{Schwartz2011}
Schwartz,~T.; Hutchison,~J.~A.; Genet,~C.; Ebbesen,~T.~W. {Reversible Switching
  of Ultrastrong Light-Molecule Coupling}. \emph{Physical Review Letters}
  \textbf{2011}, \emph{106}, 196405\relax
\mciteBstWouldAddEndPuncttrue
\mciteSetBstMidEndSepPunct{\mcitedefaultmidpunct}
{\mcitedefaultendpunct}{\mcitedefaultseppunct}\relax
\EndOfBibitem
\bibitem[K{\'{e}}na-Cohen \latin{et~al.}(2013)K{\'{e}}na-Cohen, Maier, and
  Bradley]{Kena-Cohen2013}
K{\'{e}}na-Cohen,~S.; Maier,~S.~A.; Bradley,~D. D.~C. {Ultrastrongly Coupled
  Exciton-Polaritons in Metal-Clad Organic Semiconductor Microcavities}.
  \emph{Advanced Optical Materials} \textbf{2013}, \emph{1}, 827--833\relax
\mciteBstWouldAddEndPuncttrue
\mciteSetBstMidEndSepPunct{\mcitedefaultmidpunct}
{\mcitedefaultendpunct}{\mcitedefaultseppunct}\relax
\EndOfBibitem
\bibitem[Gambino \latin{et~al.}(2014)Gambino, Mazzeo, Genco, {Di Stefano},
  Savasta, Patan{\`{e}}, Ballarini, Mangione, Lerario, Sanvitto, and
  Gigli]{Gambino2014}
Gambino,~S.; Mazzeo,~M.; Genco,~A.; {Di Stefano},~O.; Savasta,~S.;
  Patan{\`{e}},~S.; Ballarini,~D.; Mangione,~F.; Lerario,~G.; Sanvitto,~D.;
  Gigli,~G. {Exploring Light–Matter Interaction Phenomena under Ultrastrong
  Coupling Regime}. \emph{ACS Photonics} \textbf{2014}, \emph{1},
  1042--1048\relax
\mciteBstWouldAddEndPuncttrue
\mciteSetBstMidEndSepPunct{\mcitedefaultmidpunct}
{\mcitedefaultendpunct}{\mcitedefaultseppunct}\relax
\EndOfBibitem
\bibitem[Mazzeo \latin{et~al.}(2014)Mazzeo, Genco, Gambino, Ballarini,
  Mangione, {Di Stefano}, Patan{\`{e}}, Savasta, Sanvitto, and
  Gigli]{Mazzeo2014}
Mazzeo,~M.; Genco,~A.; Gambino,~S.; Ballarini,~D.; Mangione,~F.; {Di
  Stefano},~O.; Patan{\`{e}},~S.; Savasta,~S.; Sanvitto,~D.; Gigli,~G.
  {Ultrastrong light-matter coupling in electrically doped microcavity organic
  light emitting diodes}. \emph{Applied Physics Letters} \textbf{2014},
  \emph{104}, 233303\relax
\mciteBstWouldAddEndPuncttrue
\mciteSetBstMidEndSepPunct{\mcitedefaultmidpunct}
{\mcitedefaultendpunct}{\mcitedefaultseppunct}\relax
\EndOfBibitem
\bibitem[Suzuki \latin{et~al.}(2019)Suzuki, Nishiyama, Kani, Yu, Uzumi,
  Funahashi, Shimokawa, Nakanishi, and Tsurumachi]{Suzuki2019}
Suzuki,~M.; Nishiyama,~K.; Kani,~N.; Yu,~X.; Uzumi,~K.; Funahashi,~M.;
  Shimokawa,~F.; Nakanishi,~S.; Tsurumachi,~N. {Observation of
  ultrastrong-coupling regime in the Fabry–P{\'{e}}rot microcavities made of
  metal mirrors containing Lemke dye}. \emph{Applied Physics Letters}
  \textbf{2019}, \emph{114}, 191108\relax
\mciteBstWouldAddEndPuncttrue
\mciteSetBstMidEndSepPunct{\mcitedefaultmidpunct}
{\mcitedefaultendpunct}{\mcitedefaultseppunct}\relax
\EndOfBibitem
\bibitem[Liu \latin{et~al.}(2015)Liu, Rai, Grezmak, Twieg, and Singer]{Liu2015}
Liu,~B.; Rai,~P.; Grezmak,~J.; Twieg,~R.~J.; Singer,~K.~D. {Coupling of
  exciton-polaritons in low − Q coupled microcavities beyond the rotating
  wave approximation}. \emph{Physical Review B} \textbf{2015}, \emph{92},
  155301\relax
\mciteBstWouldAddEndPuncttrue
\mciteSetBstMidEndSepPunct{\mcitedefaultmidpunct}
{\mcitedefaultendpunct}{\mcitedefaultseppunct}\relax
\EndOfBibitem
\bibitem[Liu \latin{et~al.}(2019)Liu, Crescimanno, Twieg, and Singer]{Liu2019}
Liu,~B.; Crescimanno,~M.; Twieg,~R.~J.; Singer,~K.~D. {Dispersion of
  Third‐Harmonic Generation in Organic Cavity Polaritons}. \emph{Advanced
  Optical Materials} \textbf{2019}, 1801682\relax
\mciteBstWouldAddEndPuncttrue
\mciteSetBstMidEndSepPunct{\mcitedefaultmidpunct}
{\mcitedefaultendpunct}{\mcitedefaultseppunct}\relax
\EndOfBibitem
\bibitem[{Frisk Kockum} \latin{et~al.}(2019){Frisk Kockum}, Miranowicz, {De
  Liberato}, Savasta, and Nori]{FriskKockum2019}
{Frisk Kockum},~A.; Miranowicz,~A.; {De Liberato},~S.; Savasta,~S.; Nori,~F.
  {Ultrastrong coupling between light and matter}. \emph{Nature Reviews
  Physics} \textbf{2019}, \emph{1}, 19--40\relax
\mciteBstWouldAddEndPuncttrue
\mciteSetBstMidEndSepPunct{\mcitedefaultmidpunct}
{\mcitedefaultendpunct}{\mcitedefaultseppunct}\relax
\EndOfBibitem
\bibitem[Forn-D{\'{i}}az \latin{et~al.}(2019)Forn-D{\'{i}}az, Lamata, Rico,
  Kono, and Solano]{Forn-Diaz2019}
Forn-D{\'{i}}az,~P.; Lamata,~L.; Rico,~E.; Kono,~J.; Solano,~E. {Ultrastrong
  coupling regimes of light-matter interaction}. \emph{Reviews of Modern
  Physics} \textbf{2019}, \emph{91}, 025005\relax
\mciteBstWouldAddEndPuncttrue
\mciteSetBstMidEndSepPunct{\mcitedefaultmidpunct}
{\mcitedefaultendpunct}{\mcitedefaultseppunct}\relax
\EndOfBibitem
\bibitem[Askenazi \latin{et~al.}(2017)Askenazi, Vasanelli, Todorov, Sakat,
  Greffet, Beaudoin, Sagnes, and Sirtori]{Askenazi2017}
Askenazi,~B.; Vasanelli,~A.; Todorov,~Y.; Sakat,~E.; Greffet,~J.-J.;
  Beaudoin,~G.; Sagnes,~I.; Sirtori,~C. {Midinfrared Ultrastrong Light–Matter
  Coupling for THz Thermal Emission}. \emph{ACS Photonics} \textbf{2017},
  \emph{4}, 2550--2555\relax
\mciteBstWouldAddEndPuncttrue
\mciteSetBstMidEndSepPunct{\mcitedefaultmidpunct}
{\mcitedefaultendpunct}{\mcitedefaultseppunct}\relax
\EndOfBibitem
\bibitem[Yoshihara \latin{et~al.}(2018)Yoshihara, Fuse, Ao, Ashhab, Kakuyanagi,
  Saito, Aoki, Koshino, and Semba]{Yoshihara2018}
Yoshihara,~F.; Fuse,~T.; Ao,~Z.; Ashhab,~S.; Kakuyanagi,~K.; Saito,~S.;
  Aoki,~T.; Koshino,~K.; Semba,~K. {Inversion of Qubit Energy Levels in
  Qubit-Oscillator Circuits in the Deep-Strong-Coupling Regime}. \emph{Physical
  Review Letters} \textbf{2018}, \emph{120}, 183601\relax
\mciteBstWouldAddEndPuncttrue
\mciteSetBstMidEndSepPunct{\mcitedefaultmidpunct}
{\mcitedefaultendpunct}{\mcitedefaultseppunct}\relax
\EndOfBibitem
\bibitem[Bayer \latin{et~al.}(2017)Bayer, Pozimski, Schambeck, Schuh, Huber,
  Bougeard, and Lange]{Bayer2017}
Bayer,~A.; Pozimski,~M.; Schambeck,~S.; Schuh,~D.; Huber,~R.; Bougeard,~D.;
  Lange,~C. {Terahertz Light–Matter Interaction beyond Unity Coupling
  Strength}. \emph{Nano Letters} \textbf{2017}, \emph{17}, 6340--6344\relax
\mciteBstWouldAddEndPuncttrue
\mciteSetBstMidEndSepPunct{\mcitedefaultmidpunct}
{\mcitedefaultendpunct}{\mcitedefaultseppunct}\relax
\EndOfBibitem
\bibitem[Benz \latin{et~al.}(2016)Benz, Schmidt, Dreismann, Chikkaraddy, Zhang,
  Demetriadou, Carnegie, Ohadi, de~Nijs, Esteban, Aizpurua, and
  Baumberg]{Benz2016}
Benz,~F.; Schmidt,~M.~K.; Dreismann,~A.; Chikkaraddy,~R.; Zhang,~Y.;
  Demetriadou,~A.; Carnegie,~C.; Ohadi,~H.; de~Nijs,~B.; Esteban,~R.;
  Aizpurua,~J.; Baumberg,~J.~J. {Single-molecule optomechanics in
  "picocavities".} \emph{Science (New York, N.Y.)} \textbf{2016}, \emph{354},
  726--729\relax
\mciteBstWouldAddEndPuncttrue
\mciteSetBstMidEndSepPunct{\mcitedefaultmidpunct}
{\mcitedefaultendpunct}{\mcitedefaultseppunct}\relax
\EndOfBibitem
\bibitem[Ciuti \latin{et~al.}(2005)Ciuti, Bastard, and Carusotto]{Ciuti2005}
Ciuti,~C.; Bastard,~G.; Carusotto,~I. {Quantum vacuum properties of the
  intersubband cavity polariton field}. \emph{Physical Review B} \textbf{2005},
  \emph{72}, 115303\relax
\mciteBstWouldAddEndPuncttrue
\mciteSetBstMidEndSepPunct{\mcitedefaultmidpunct}
{\mcitedefaultendpunct}{\mcitedefaultseppunct}\relax
\EndOfBibitem
\bibitem[George \latin{et~al.}(2015)George, Wang, Chervy, Canaguier-Durand,
  Schaeffer, Lehn, Hutchison, Genet, and Ebbesen]{George2015}
George,~J.; Wang,~S.; Chervy,~T.; Canaguier-Durand,~A.; Schaeffer,~G.;
  Lehn,~J.-M.; Hutchison,~J.~A.; Genet,~C.; Ebbesen,~T.~W. {Ultra-strong
  coupling of molecular materials: spectroscopy and dynamics}. \emph{Faraday
  Discussions} \textbf{2015}, \emph{178}, 281--294\relax
\mciteBstWouldAddEndPuncttrue
\mciteSetBstMidEndSepPunct{\mcitedefaultmidpunct}
{\mcitedefaultendpunct}{\mcitedefaultseppunct}\relax
\EndOfBibitem
\bibitem[Tropf and Gather(2018)Tropf, and Gather]{Tropf2018}
Tropf,~L.; Gather,~M.~C. {Investigating the Onset of the Strong Coupling Regime
  by Fine-Tuning the Rabi Splitting in Multilayer Organic Microcavities}.
  \emph{Advanced Optical Materials} \textbf{2018}, \emph{6}, 1800203\relax
\mciteBstWouldAddEndPuncttrue
\mciteSetBstMidEndSepPunct{\mcitedefaultmidpunct}
{\mcitedefaultendpunct}{\mcitedefaultseppunct}\relax
\EndOfBibitem
\bibitem[Barachati \latin{et~al.}(2018)Barachati, Simon, Getmanenko, Barlow,
  Marder, and K{\'{e}}na-Cohen]{Barachati2018}
Barachati,~F.; Simon,~J.; Getmanenko,~Y.~A.; Barlow,~S.; Marder,~S.~R.;
  K{\'{e}}na-Cohen,~S. {Tunable Third-Harmonic Generation from Polaritons in
  the Ultrastrong Coupling Regime}. \emph{ACS Photonics} \textbf{2018},
  \emph{5}, 119--125\relax
\mciteBstWouldAddEndPuncttrue
\mciteSetBstMidEndSepPunct{\mcitedefaultmidpunct}
{\mcitedefaultendpunct}{\mcitedefaultseppunct}\relax
\EndOfBibitem
\bibitem[Campoy-Quiles \latin{et~al.}(2008)Campoy-Quiles, Ferenczi,
  Agostinelli, Etchegoin, Kim, Anthopoulos, Stavrinou, Bradley, and
  Nelson]{Campoy-Quiles2008}
Campoy-Quiles,~M.; Ferenczi,~T.; Agostinelli,~T.; Etchegoin,~P.~G.; Kim,~Y.;
  Anthopoulos,~T.~D.; Stavrinou,~P.~N.; Bradley,~D. D.~C.; Nelson,~J.
  {Morphology evolution via self-organization and lateral and vertical
  diffusion in polymer:fullerene solar cell blends}. \emph{Nature Materials}
  \textbf{2008}, \emph{7}, 158--164\relax
\mciteBstWouldAddEndPuncttrue
\mciteSetBstMidEndSepPunct{\mcitedefaultmidpunct}
{\mcitedefaultendpunct}{\mcitedefaultseppunct}\relax
\EndOfBibitem
\bibitem[{Le Roux} and Bradley(2018){Le Roux}, and Bradley]{LeRoux2018}
{Le Roux},~F.; Bradley,~D. D.~C. {Conformational control of exciton-polariton
  physics in metal-poly(9,9-dioctylfluorene)-metal cavities}. \emph{Physical
  Review B} \textbf{2018}, \emph{98}, 195306\relax
\mciteBstWouldAddEndPuncttrue
\mciteSetBstMidEndSepPunct{\mcitedefaultmidpunct}
{\mcitedefaultendpunct}{\mcitedefaultseppunct}\relax
\EndOfBibitem
\bibitem[Perevedentsev \latin{et~al.}(2016)Perevedentsev, Chander, Kim, and
  Bradley]{Perevedentsev2016}
Perevedentsev,~A.; Chander,~N.; Kim,~J.-S.; Bradley,~D. D.~C. {Spectroscopic
  properties of poly(9,9-dioctylfluorene) thin films possessing varied
  fractions of $\beta$-phase chain segments: enhanced photoluminescence
  efficiency via conformation structuring}. \emph{Journal of Polymer Science
  Part B: Polymer Physics} \textbf{2016}, \emph{54}, 1995--2006\relax
\mciteBstWouldAddEndPuncttrue
\mciteSetBstMidEndSepPunct{\mcitedefaultmidpunct}
{\mcitedefaultendpunct}{\mcitedefaultseppunct}\relax
\EndOfBibitem
\bibitem[Virgili \latin{et~al.}(2001)Virgili, Lidzey, Grell, Walker, Asimakis,
  and Bradley]{Virgili2001}
Virgili,~T.; Lidzey,~D.; Grell,~M.; Walker,~S.; Asimakis,~A.; Bradley,~D.
  {Completely polarized photoluminescence emission from a microcavity
  containing an aligned conjugated polymer}. \emph{Chemical Physics Letters}
  \textbf{2001}, \emph{341}, 219--224\relax
\mciteBstWouldAddEndPuncttrue
\mciteSetBstMidEndSepPunct{\mcitedefaultmidpunct}
{\mcitedefaultendpunct}{\mcitedefaultseppunct}\relax
\EndOfBibitem
\bibitem[Campoy-Quiles \latin{et~al.}(2005)Campoy-Quiles, Etchegoin, and
  Bradley]{Campoy-Quiles2005b}
Campoy-Quiles,~M.; Etchegoin,~P.~G.; Bradley,~D. D.~C. {On the optical
  anisotropy of conjugated polymer thin films}. \emph{Physical Review B}
  \textbf{2005}, \emph{72}, 045209\relax
\mciteBstWouldAddEndPuncttrue
\mciteSetBstMidEndSepPunct{\mcitedefaultmidpunct}
{\mcitedefaultendpunct}{\mcitedefaultseppunct}\relax
\EndOfBibitem
\bibitem[Hertzog \latin{et~al.}(2017)Hertzog, Rudquist, Hutchison, George,
  Ebbesen, and B{\"{o}}rjesson]{Hertzog2017}
Hertzog,~M.; Rudquist,~P.; Hutchison,~J.~A.; George,~J.; Ebbesen,~T.~W.;
  B{\"{o}}rjesson,~K. {Voltage-Controlled Switching of Strong Light-Matter
  Interactions using Liquid Crystals}. \emph{Chemistry - A European Journal}
  \textbf{2017}, \emph{23}, 18166--18170\relax
\mciteBstWouldAddEndPuncttrue
\mciteSetBstMidEndSepPunct{\mcitedefaultmidpunct}
{\mcitedefaultendpunct}{\mcitedefaultseppunct}\relax
\EndOfBibitem
\bibitem[Gao \latin{et~al.}(2018)Gao, Li, Bamba, and Kono]{Gao2018}
Gao,~W.; Li,~X.; Bamba,~M.; Kono,~J. {Continuous transition between weak and
  ultrastrong coupling through exceptional points in carbon nanotube
  microcavity exciton–polaritons}. \emph{Nature Photonics} \textbf{2018},
  \emph{12}, 362--367\relax
\mciteBstWouldAddEndPuncttrue
\mciteSetBstMidEndSepPunct{\mcitedefaultmidpunct}
{\mcitedefaultendpunct}{\mcitedefaultseppunct}\relax
\EndOfBibitem
\bibitem[Grell \latin{et~al.}(1997)Grell, Bradley, Inbasekaran, and
  Woo]{Grell1997}
Grell,~M.; Bradley,~D. D.~C.; Inbasekaran,~M.; Woo,~E.~P. {A glass-forming
  conjugated main-chain liquid crystal polymer for polarized
  electroluminescence applications}. \emph{Advanced Materials} \textbf{1997},
  \emph{9}, 798--802\relax
\mciteBstWouldAddEndPuncttrue
\mciteSetBstMidEndSepPunct{\mcitedefaultmidpunct}
{\mcitedefaultendpunct}{\mcitedefaultseppunct}\relax
\EndOfBibitem
\bibitem[Grell \latin{et~al.}(1999)Grell, Redecker, Whitehead, Bradley,
  Inbasekaran, and Woo]{GRELL1999}
Grell,~M.; Redecker,~M.; Whitehead,~K.~S.; Bradley,~D. D.~C.; Inbasekaran,~M.;
  Woo,~E.~P. {Monodomain alignment of thermotropic fluorene copolymers}.
  \emph{Liquid Crystals} \textbf{1999}, \emph{26}, 1403--1407\relax
\mciteBstWouldAddEndPuncttrue
\mciteSetBstMidEndSepPunct{\mcitedefaultmidpunct}
{\mcitedefaultendpunct}{\mcitedefaultseppunct}\relax
\EndOfBibitem
\bibitem[Whitehead \latin{et~al.}(2000)Whitehead, Grell, Bradley, Jandke, and
  Strohriegl]{Whitehead2000a}
Whitehead,~K.~S.; Grell,~M.; Bradley,~D. D.~C.; Jandke,~M.; Strohriegl,~P.
  {Highly polarized blue electroluminescence from homogeneously aligned films
  of poly(9,9-dioctylfluorene)}. \emph{Applied Physics Letters} \textbf{2000},
  \emph{76}, 2946--2948\relax
\mciteBstWouldAddEndPuncttrue
\mciteSetBstMidEndSepPunct{\mcitedefaultmidpunct}
{\mcitedefaultendpunct}{\mcitedefaultseppunct}\relax
\EndOfBibitem
\bibitem[Dyreklev \latin{et~al.}(1995)Dyreklev, Berggren, Ingan{\"{a}}s,
  Andersson, Wennerstr{\"{o}}m, and Hjertberg]{Dyreklev1995}
Dyreklev,~P.; Berggren,~M.; Ingan{\"{a}}s,~O.; Andersson,~M.~R.;
  Wennerstr{\"{o}}m,~O.; Hjertberg,~T. {Polarized electroluminescence from an
  oriented substituted polythiophene in a light emitting diode}. \emph{Advanced
  Materials} \textbf{1995}, \emph{7}, 43--45\relax
\mciteBstWouldAddEndPuncttrue
\mciteSetBstMidEndSepPunct{\mcitedefaultmidpunct}
{\mcitedefaultendpunct}{\mcitedefaultseppunct}\relax
\EndOfBibitem
\bibitem[Cimrov{\'{a}} \latin{et~al.}(1996)Cimrov{\'{a}}, Remmers, Neher, and
  Wegner]{Cimrova1996}
Cimrov{\'{a}},~V.; Remmers,~M.; Neher,~D.; Wegner,~G. {Polarized light emission
  from LEDs prepared by the Langmuir-Blodgett technique}. \emph{Advanced
  Materials} \textbf{1996}, \emph{8}, 146--149\relax
\mciteBstWouldAddEndPuncttrue
\mciteSetBstMidEndSepPunct{\mcitedefaultmidpunct}
{\mcitedefaultendpunct}{\mcitedefaultseppunct}\relax
\EndOfBibitem
\bibitem[Hagler \latin{et~al.}(1991)Hagler, Pakbaz, Voss, and
  Heeger]{Hagler1991}
Hagler,~T.~W.; Pakbaz,~K.; Voss,~K.~F.; Heeger,~A.~J. {Enhanced order and
  electronic delocalization in conjugated polymers oriented by gel processing
  in polyethylene}. \emph{Physical Review B} \textbf{1991}, \emph{44},
  8652--8666\relax
\mciteBstWouldAddEndPuncttrue
\mciteSetBstMidEndSepPunct{\mcitedefaultmidpunct}
{\mcitedefaultendpunct}{\mcitedefaultseppunct}\relax
\EndOfBibitem
\bibitem[Zheng \latin{et~al.}(2007)Zheng, Yim, Saifullah, Welland, Friend, Kim,
  and S.]{ZijianZheng2007}
Zheng,~Z.; Yim,~K.; Saifullah,~M.; Welland,~M.~E.; Friend,~R.~H.; Kim,~J.-S.;
  S.,~H. W.~T. {Uniaxial Alignment of Liquid-Crystalline Conjugated Polymers by
  Nanoconfinement}. \textbf{2007}, \relax
\mciteBstWouldAddEndPunctfalse
\mciteSetBstMidEndSepPunct{\mcitedefaultmidpunct}
{}{\mcitedefaultseppunct}\relax
\EndOfBibitem
\bibitem[Whitehead \latin{et~al.}(2000)Whitehead, Grell, Bradley, Inbasekaran,
  and Woo]{Whitehead2000}
Whitehead,~K.; Grell,~M.; Bradley,~D.; Inbasekaran,~M.; Woo,~E. {Polarized
  emission from liquid crystal polymers}. \emph{Synthetic Metals}
  \textbf{2000}, \emph{111-112}, 181--185\relax
\mciteBstWouldAddEndPuncttrue
\mciteSetBstMidEndSepPunct{\mcitedefaultmidpunct}
{\mcitedefaultendpunct}{\mcitedefaultseppunct}\relax
\EndOfBibitem
\bibitem[Schmid \latin{et~al.}(2008)Schmid, Yim, Chang, Zheng, Huck, Friend,
  Kim, and Herz]{Schmid2008}
Schmid,~S.~A.; Yim,~K.~H.; Chang,~M.~H.; Zheng,~Z.; Huck,~W. T.~S.;
  Friend,~R.~H.; Kim,~J.~S.; Herz,~L.~M. {Polarization anisotropy dynamics for
  thin films of a conjugated polymer aligned by nanoimprinting}. \emph{Physical
  Review B} \textbf{2008}, \emph{77}, 115338\relax
\mciteBstWouldAddEndPuncttrue
\mciteSetBstMidEndSepPunct{\mcitedefaultmidpunct}
{\mcitedefaultendpunct}{\mcitedefaultseppunct}\relax
\EndOfBibitem
\bibitem[Lee \latin{et~al.}(2017)Lee, Lee, Kim, and Yu]{Lee2017}
Lee,~D.-M.; Lee,~Y.-J.; Kim,~J.-H.; Yu,~C.-J. {Birefringence-dependent
  linearly-polarized emission in a liquid crystalline organic light emitting
  polymer}. \emph{Optics Express} \textbf{2017}, \emph{25}, 3737\relax
\mciteBstWouldAddEndPuncttrue
\mciteSetBstMidEndSepPunct{\mcitedefaultmidpunct}
{\mcitedefaultendpunct}{\mcitedefaultseppunct}\relax
\EndOfBibitem
\bibitem[Misaki \latin{et~al.}(2008)Misaki, Chikamatsu, Yoshida, Azumi,
  Tanigaki, Yase, Nagamatsu, and Ueda]{Misaki2008}
Misaki,~M.; Chikamatsu,~M.; Yoshida,~Y.; Azumi,~R.; Tanigaki,~N.; Yase,~K.;
  Nagamatsu,~S.; Ueda,~Y. {Highly efficient polarized polymer light-emitting
  diodes utilizing oriented films of $\beta$-phase poly(9,9-dioctylfluorene)}.
  \emph{Applied Physics Letters} \textbf{2008}, \emph{93}, 023304\relax
\mciteBstWouldAddEndPuncttrue
\mciteSetBstMidEndSepPunct{\mcitedefaultmidpunct}
{\mcitedefaultendpunct}{\mcitedefaultseppunct}\relax
\EndOfBibitem
\bibitem[Sirringhaus \latin{et~al.}(2000)Sirringhaus, Wilson, Friend,
  Inbasekaran, Wu, Woo, Grell, and Bradley]{Sirringhaus2000}
Sirringhaus,~H.; Wilson,~R.~J.; Friend,~R.~H.; Inbasekaran,~M.; Wu,~W.;
  Woo,~E.~P.; Grell,~M.; Bradley,~D. D.~C. {Mobility enhancement in conjugated
  polymer field-effect transistors through chain alignment in a
  liquid-crystalline phase}. \emph{Applied Physics Letters} \textbf{2000},
  \emph{77}, 406--408\relax
\mciteBstWouldAddEndPuncttrue
\mciteSetBstMidEndSepPunct{\mcitedefaultmidpunct}
{\mcitedefaultendpunct}{\mcitedefaultseppunct}\relax
\EndOfBibitem
\bibitem[Seki(2014)]{Seki2014}
Seki,~T. {New strategies and implications for the photoalignment of liquid
  crystalline polymers}. \emph{Polymer Journal} \textbf{2014}, \emph{46},
  751--768\relax
\mciteBstWouldAddEndPuncttrue
\mciteSetBstMidEndSepPunct{\mcitedefaultmidpunct}
{\mcitedefaultendpunct}{\mcitedefaultseppunct}\relax
\EndOfBibitem
\bibitem[Li \latin{et~al.}(2006)Li, Kozenkov, Yeung, Xu, Chigrinov, and
  Kwok]{Li2006}
Li,~X.; Kozenkov,~V.~M.; Yeung,~F. S.-Y.; Xu,~P.; Chigrinov,~V.~G.; Kwok,~H.-S.
  {Liquid-Crystal Photoalignment by Super Thin Azo Dye Layer}. \emph{Japanese
  Journal of Applied Physics} \textbf{2006}, \emph{45}, 203--205\relax
\mciteBstWouldAddEndPuncttrue
\mciteSetBstMidEndSepPunct{\mcitedefaultmidpunct}
{\mcitedefaultendpunct}{\mcitedefaultseppunct}\relax
\EndOfBibitem
\bibitem[Ma \latin{et~al.}(2015)Ma, Li, Li, Ji, Luo, Zheng, Cai, Chigrinov, Lu,
  Hu, and Chen]{Ma2015}
Ma,~L.-L.; Li,~S.-S.; Li,~W.-S.; Ji,~W.; Luo,~B.; Zheng,~Z.-G.; Cai,~Z.-P.;
  Chigrinov,~V.; Lu,~Y.-Q.; Hu,~W.; Chen,~L.-J. {Rationally Designed Dynamic
  Superstructures Enabled by Photoaligning Cholesteric Liquid Crystals}.
  \emph{Advanced Optical Materials} \textbf{2015}, \emph{3}, 1691--1696\relax
\mciteBstWouldAddEndPuncttrue
\mciteSetBstMidEndSepPunct{\mcitedefaultmidpunct}
{\mcitedefaultendpunct}{\mcitedefaultseppunct}\relax
\EndOfBibitem
\bibitem[Zhang \latin{et~al.}(2019)Zhang, Ma, Zhang, Shi, Fang, and
  Bradley]{Zhang2019}
Zhang,~H.; Ma,~L.; Zhang,~Q.; Shi,~Y.; Fang,~Y.; Bradley,~D. D.~C. \emph{In
  Preparation} \textbf{2019}, \relax
\mciteBstWouldAddEndPunctfalse
\mciteSetBstMidEndSepPunct{\mcitedefaultmidpunct}
{}{\mcitedefaultseppunct}\relax
\EndOfBibitem
\bibitem[Ma \latin{et~al.}(2017)Ma, Tang, Hu, Cui, Ge, Chen, Chen, Qian, Chi,
  and Lu]{Ma2017}
Ma,~L.-L.; Tang,~M.-J.; Hu,~W.; Cui,~Z.-Q.; Ge,~S.-J.; Chen,~P.; Chen,~L.-J.;
  Qian,~H.; Chi,~L.-F.; Lu,~Y.-Q. {Smectic Layer Origami via Preprogrammed
  Photoalignment}. \emph{Advanced Materials} \textbf{2017}, \emph{29},
  1606671\relax
\mciteBstWouldAddEndPuncttrue
\mciteSetBstMidEndSepPunct{\mcitedefaultmidpunct}
{\mcitedefaultendpunct}{\mcitedefaultseppunct}\relax
\EndOfBibitem
\bibitem[Sanvitto and K{\'{e}}na-Cohen(2016)Sanvitto, and
  K{\'{e}}na-Cohen]{Sanvitto2016}
Sanvitto,~D.; K{\'{e}}na-Cohen,~S. {The road towards polaritonic devices}.
  \emph{Nature Materials} \textbf{2016}, \emph{15}, 1061--1073\relax
\mciteBstWouldAddEndPuncttrue
\mciteSetBstMidEndSepPunct{\mcitedefaultmidpunct}
{\mcitedefaultendpunct}{\mcitedefaultseppunct}\relax
\EndOfBibitem
\bibitem[Solnyshkov \latin{et~al.}(2015)Solnyshkov, Bleu, and
  Malpuech]{Solnyshkov2015}
Solnyshkov,~D.; Bleu,~O.; Malpuech,~G. {All optical controlled-NOT gate based
  on an exciton–polariton circuit}. \emph{Superlattices and Microstructures}
  \textbf{2015}, \emph{83}, 466--475\relax
\mciteBstWouldAddEndPuncttrue
\mciteSetBstMidEndSepPunct{\mcitedefaultmidpunct}
{\mcitedefaultendpunct}{\mcitedefaultseppunct}\relax
\EndOfBibitem
\bibitem[Espinosa-Ortega and Liew(2013)Espinosa-Ortega, and
  Liew]{Espinosa-Ortega2013}
Espinosa-Ortega,~T.; Liew,~T. C.~H. {Complete architecture of integrated
  photonic circuits based on and and not logic gates of exciton polaritons in
  semiconductor microcavities}. \emph{Physical Review B} \textbf{2013},
  \emph{87}, 195305\relax
\mciteBstWouldAddEndPuncttrue
\mciteSetBstMidEndSepPunct{\mcitedefaultmidpunct}
{\mcitedefaultendpunct}{\mcitedefaultseppunct}\relax
\EndOfBibitem
\bibitem[Campoy-Quiles \latin{et~al.}(2005)Campoy-Quiles, Heliotis, Xia, Ariu,
  Pintani, Etchegoin, and Bradley]{Campoy-Quiles2005a}
Campoy-Quiles,~M.; Heliotis,~G.; Xia,~R.; Ariu,~M.; Pintani,~M.; Etchegoin,~P.;
  Bradley,~D. D.~C. {Ellipsometric Characterization of the Optical Constants of
  Polyfluorene Gain Media}. \emph{Advanced Functional Materials} \textbf{2005},
  \emph{15}, 925--933\relax
\mciteBstWouldAddEndPuncttrue
\mciteSetBstMidEndSepPunct{\mcitedefaultmidpunct}
{\mcitedefaultendpunct}{\mcitedefaultseppunct}\relax
\EndOfBibitem
\bibitem[Campoy-Quiles \latin{et~al.}(2014)Campoy-Quiles, Alonso, Bradley, and
  Richter]{Campoy-Quiles2014}
Campoy-Quiles,~M.; Alonso,~M.~I.; Bradley,~D. D.~C.; Richter,~L.~J. {Advanced
  Ellipsometric Characterization of Conjugated Polymer Films}. \emph{Advanced
  Functional Materials} \textbf{2014}, \emph{24}, 2116--2134\relax
\mciteBstWouldAddEndPuncttrue
\mciteSetBstMidEndSepPunct{\mcitedefaultmidpunct}
{\mcitedefaultendpunct}{\mcitedefaultseppunct}\relax
\EndOfBibitem
\bibitem[Campoy-Quiles \latin{et~al.}(2005)Campoy-Quiles, Etchegoin, and
  Bradley]{Campoy-Quiles2005}
Campoy-Quiles,~M.; Etchegoin,~P.~G.; Bradley,~D. D.~C. {On the optical
  anisotropy of conjugated polymer thin films}. \emph{Physical Review B}
  \textbf{2005}, \emph{72}, 045209\relax
\mciteBstWouldAddEndPuncttrue
\mciteSetBstMidEndSepPunct{\mcitedefaultmidpunct}
{\mcitedefaultendpunct}{\mcitedefaultseppunct}\relax
\EndOfBibitem
\bibitem[Tropf \latin{et~al.}(2017)Tropf, Dietrich, Herbst, Kanibolotsky,
  Skabara, W{\"{u}}rthner, Samuel, Gather, and H{\"{o}}fling]{Tropf2017}
Tropf,~L.; Dietrich,~C.~P.; Herbst,~S.; Kanibolotsky,~A.~L.; Skabara,~P.~J.;
  W{\"{u}}rthner,~F.; Samuel,~I. D.~W.; Gather,~M.~C.; H{\"{o}}fling,~S.
  {Influence of optical material properties on strong coupling in organic
  semiconductor based microcavities}. \emph{Applied Physics Letters}
  \textbf{2017}, \emph{110}, 153302\relax
\mciteBstWouldAddEndPuncttrue
\mciteSetBstMidEndSepPunct{\mcitedefaultmidpunct}
{\mcitedefaultendpunct}{\mcitedefaultseppunct}\relax
\EndOfBibitem
\bibitem[Rothe \latin{et~al.}(2006)Rothe, Galbrecht, Scherf, and
  Monkman]{Rothe2006}
Rothe,~C.; Galbrecht,~F.; Scherf,~U.; Monkman,~A. {The $\beta$-Phase of
  Poly(9,9-dioctylfluorene) as a Potential System for Electrically Pumped
  Organic Lasing}. \emph{Advanced Materials} \textbf{2006}, \emph{18},
  2137--2140\relax
\mciteBstWouldAddEndPuncttrue
\mciteSetBstMidEndSepPunct{\mcitedefaultmidpunct}
{\mcitedefaultendpunct}{\mcitedefaultseppunct}\relax
\EndOfBibitem
\bibitem[Valyukh \latin{et~al.}(2008)Valyukh, Arwin, Chigrinov, and
  Valyukh]{Valyukh2008}
Valyukh,~I.; Arwin,~H.; Chigrinov,~V.; Valyukh,~S. {UV-induced in-plane
  anisotropy in layers of mixture of the azo-dyes SD-1/SDA-2 characterized by
  spectroscopic ellipsometry}. \emph{physica status solidi (c)} \textbf{2008},
  \emph{5}, 1274--1277\relax
\mciteBstWouldAddEndPuncttrue
\mciteSetBstMidEndSepPunct{\mcitedefaultmidpunct}
{\mcitedefaultendpunct}{\mcitedefaultseppunct}\relax
\EndOfBibitem
\bibitem[Agranovich(1957)]{Agranovich1957}
Agranovich,~V.~M. \emph{Opt Spektrosk.} \textbf{1957}, \emph{2}\relax
\mciteBstWouldAddEndPuncttrue
\mciteSetBstMidEndSepPunct{\mcitedefaultmidpunct}
{\mcitedefaultendpunct}{\mcitedefaultseppunct}\relax
\EndOfBibitem
\bibitem[Hopfield(1958)]{Hopfield1958}
Hopfield,~J.~J. {Theory of the Contribution of Excitons to the Complex
  Dielectric Constant of Crystals}. \emph{Physical Review} \textbf{1958},
  \emph{112}, 1555--1567\relax
\mciteBstWouldAddEndPuncttrue
\mciteSetBstMidEndSepPunct{\mcitedefaultmidpunct}
{\mcitedefaultendpunct}{\mcitedefaultseppunct}\relax
\EndOfBibitem
\bibitem[Economou(1969)]{Economou1969}
Economou,~E.~N. {Surface Plasmons in Thin Films}. \emph{Physical Review}
  \textbf{1969}, \emph{182}, 539--554\relax
\mciteBstWouldAddEndPuncttrue
\mciteSetBstMidEndSepPunct{\mcitedefaultmidpunct}
{\mcitedefaultendpunct}{\mcitedefaultseppunct}\relax
\EndOfBibitem
\bibitem[Litinskaya and Agranovich(2012)Litinskaya, and
  Agranovich]{Litinskaya2012}
Litinskaya,~M.; Agranovich,~V.~M. {Polariton trap in microcavities with
  metallic mirrors}. \emph{Journal of Physics: Condensed Matter} \textbf{2012},
  \emph{24}, 015302\relax
\mciteBstWouldAddEndPuncttrue
\mciteSetBstMidEndSepPunct{\mcitedefaultmidpunct}
{\mcitedefaultendpunct}{\mcitedefaultseppunct}\relax
\EndOfBibitem
\bibitem[Perevedentsev \latin{et~al.}(2015)Perevedentsev, Sonnefraud, Belton,
  Sharma, Cass, Maier, Kim, Stavrinou, and Bradley]{Perevedentsev2015}
Perevedentsev,~A.; Sonnefraud,~Y.; Belton,~C.~R.; Sharma,~S.; Cass,~A. E.~G.;
  Maier,~S.~A.; Kim,~J.-S.; Stavrinou,~P.~N.; Bradley,~D. D.~C. {Dip-pen
  patterning of poly(9,9-dioctylfluorene) chain-conformation-based
  nano-photonic elements}. \emph{Nature Communications} \textbf{2015},
  \emph{6}, 5977\relax
\mciteBstWouldAddEndPuncttrue
\mciteSetBstMidEndSepPunct{\mcitedefaultmidpunct}
{\mcitedefaultendpunct}{\mcitedefaultseppunct}\relax
\EndOfBibitem
\bibitem[Scafirimuto \latin{et~al.}(2018)Scafirimuto, Urbonas, Scherf, Mahrt,
  and St{\"{o}}ferle]{Scafirimuto2018}
Scafirimuto,~F.; Urbonas,~D.; Scherf,~U.; Mahrt,~R.~F.; St{\"{o}}ferle,~T.
  {Room-Temperature Exciton-Polariton Condensation in a Tunable
  Zero-Dimensional Microcavity}. \emph{ACS Photonics} \textbf{2018}, \emph{5},
  85--89\relax
\mciteBstWouldAddEndPuncttrue
\mciteSetBstMidEndSepPunct{\mcitedefaultmidpunct}
{\mcitedefaultendpunct}{\mcitedefaultseppunct}\relax
\EndOfBibitem
\bibitem[Rajendran \latin{et~al.}(2019)Rajendran, Wei, Ohadi, Ruseckas,
  Turnbull, and Samuel]{Rajendran2019}
Rajendran,~S.~K.; Wei,~M.; Ohadi,~H.; Ruseckas,~A.; Turnbull,~G.~A.; Samuel,~I.
  D.~W. {Low Threshold Polariton Lasing from a Solution‐Processed Organic
  Semiconductor in a Planar Microcavity}. \emph{Advanced Optical Materials}
  \textbf{2019}, \emph{7}, 1801791\relax
\mciteBstWouldAddEndPuncttrue
\mciteSetBstMidEndSepPunct{\mcitedefaultmidpunct}
{\mcitedefaultendpunct}{\mcitedefaultseppunct}\relax
\EndOfBibitem
\end{mcitethebibliography}

\end{document}